\documentclass[12pt]{article}
\usepackage{amsmath,amssymb,epsfig}
\usepackage{graphicx,subfigure,color}
\usepackage{verbatim,geometry}
\usepackage{cite}
\numberwithin{equation}{section}

\makeatletter

\def\unit{{1\kern-.65ex {\rm l}}}
\def\1{{1\kern-.65ex {\rm l}}}

\newcommand{\be}{\begin{equation}}
\newcommand{\ee}{\end{equation}}
\newcommand{\ba}{\begin{aligned}}
\newcommand{\ea}{\end{aligned}}

\def\m1{\left(-1\right)^{F_i}}

\makeatletter
\def\sla@#1#2#3#4#5{{%
  \setbox\z@\hbox{$\m@th#4#5$}%
  \setbox\tw@\hbox{$\m@th#4#1$}%
  \dimen4\wd\ifdim\wd\z@<\wd\tw@\tw@\else\z@\fi
  \dimen@\ht\tw@
  \advance\dimen@-\dp\tw@
  \advance\dimen@-\ht\z@
  \advance\dimen@\dp\z@
  \divide\dimen@\tw@
  \advance\dimen@-#3\ht\tw@
  \advance\dimen@-#3\dp\tw@
  \dimen@ii#2\wd\z@  \raise-\dimen@\hbox to\dimen4{%
    \hss\kern\dimen@ii\box\tw@\kern-\dimen@ii\hss}%
  \llap{\hbox to\dimen4{\hss\box\z@\hss}}}}
\def\slashed#1{%
  \expandafter\ifx\csname sla@\string#1\endcsname\relax
    {\mathpalette{\sla@/00}{#1}}%
  \else
    \csname sla@\string#1\endcsname
  \fi}
\makeatother

\makeatother

\newlength{\wth}
\newcommand{\twographs}[2]
{\unitlength=1.1in
\begin{picture}(5.4,2.25) 
 \put(-0.85,-0.1){\epsfig{file=#1, width=1.05 \wth}}
\put(2.25,-0.1){\epsfig{file=#2, width=1.05 \wth}}
\put(-0.1,2.35){(a)}
\put(2.9,2.35){(b)}
\end{picture}
}

\begin{document}

\numberwithin{equation}{section}  
\allowdisplaybreaks  


%
%


\thispagestyle{empty}

\vspace*{-2cm} 
\begin{center}
{\tt IPPP-11-57 \qquad  DCPT-11-114 \qquad 
EFI-11-27  \qquad 
KCL-MTH-11-17}
\end{center}

\vspace*{0.8cm} 
\begin{center}
 {\LARGE \bf  Unification and LHC Phenomenology \\ of F-theory GUTs with $U(1)_{PQ}$\\ }
\vspace*{1.1cm}
Matthew J. Dolan$^1$, Joseph Marsano$^2$ and Sakura Sch\"afer-Nameki$^3$

 \vspace*{1.0cm} 
{ \it $^1$ Institute for Particle Physics Phenomenology,\\ }
{ \it University of Durham, Durham DH1 3LE, UK\\}
{\tt m.j.dolan@durham.ac.uk\\ }

{ \it$^2$ Enrico Fermi Institute, University of Chicago\\}
{ \it 5640 S Ellis Avenue, Chicago, IL 60637 USA\\}
{\tt marsano@uchicago.edu \\}

{ \it $^3$ Department of Mathematics, King's College London \\}
{ \it The Strand, WC2R 2LS, London, UK\\}
{\tt ss299@theory.caltech.edu\\}
 \vspace*{0.8cm} 
\end{center}

\begin{center} \textbf{Abstract} \end{center}
\noindent
We undertake  a phenomenological study of SU(5) F-theory GUT models with an additional $U(1)_{PQ}$ symmetry.
In such models, breaking SU(5) with hypercharge flux leads to the presence of non-GUT multiplets in the spectrum. We study the effect these have on the unification of gauge couplings, including two-loop running as well as low- and high-scale threshold corrections. We use the requirement of unification to constrain the size of thresholds from KK modes of SU(5) gauge and matter fields. Assuming the non-GUT multiplets play the role of messengers of gauge mediation leads to controlled non-universalities in the sparticle spectrum while maintaining grand unification, and we study the LHC phenomenology of this scenario. We find that the MSSM spectrum may become compressed or stretched out {by up to a factor of three} depending on the distribution of hypercharge flux. We present a set of benchmark points whose production cross-sections and decays we investigate, and argue that precision kinematic edge measurements will allow the LHC to distinguish between our model and mGMSB.




\newpage
\tableofcontents

\newpage

\section{Introduction}

Despite major progress over the last few years in understanding conceptual aspects of F-theory realizations of $SU(5)$ GUTs\footnote{See~\cite{Heckman:2010bq,Weigand:2010wm} and references therein.},
the particle physics implications of such endeavours remain at best obscure. Expecting a robust prediction from this class of string theory models would undoubtedly at this point be aiming too high. 
However, imposing the reasonable and mild requirements of global consistency and rudimentary phenomenological soundness, it 
has recently been observed in \cite{Dolan:2011iu} that the class of $SU(5)$ GUTs that can potentially arise from complete F-theory compactifications {{can be} condensed to a surprisingly small number.  Naturally it is of  {great} interest to explore the phenomenological properties of these models, which is the subject of this paper. 

There are two main requirements that lead to the models of  \cite{Dolan:2011iu}. 
{{Firstly, unwelcome dimension 5 proton decay operators as well as a tree-level $\mu$-term are forbidden by means of a gauged $U(1)$ symmetry{\footnote{One could bypass this type of scenario altogether by using some other structure, such as discrete symmetries \cite{Hayashi:2009ge} like those recently studied in  \cite{Lee:2010gv,Lee:2011dya}, but we stick to $U(1)$'s in this paper}}.  The existence of such a symmetry, together with consistent anomaly cancellation \cite{Dudas:2010zb,Marsano:2010sq,Dolan:2011iu}, implies the presence of non-GUT exotics in the spectrum \cite{Marsano:2009gv,Marsano:2009wr}.  The second requirement is that the model be realizable in a local Calabi-Yau geometry, by which we mean that it is compatible with a spectral cover description \cite{Donagi:2009ra}.  As we saw in the survey of \cite{Dolan:2011iu}, this seems to further constrain the exotic sector.  In section \ref{sec:surveysummary}, we shall give a detailed account of the class of models that we study.
}}

{{Our subsequent analysis focuses on two aspects of the models in \cite{Dolan:2011iu}.  O}ne long-standing issue in the context of F-theory GUTs has been compatibility with precision gauge-coupling unification. In addition to standard MSSM contributions to the running of the couplings (and refinements such as NLO effects and low-scale thresholds), there are various threshold contributions that arise due to the GUT being embedded into a higher dimensional theory {{\cite{Donagi:2008kj,Blumenhagen:2008aw,Leontaris:2011pu,Leontaris:2011tw,Callaghan:2011jj}}}.  {{The exotics in our models will give new non-universal contributions that add to these and naively seem to threaten compatibility with precision unification \cite{Dolan:2011iu}}}.
{{We address this matter in section \ref{app:unification}, where we find that the requisite tuning of model-dependent high scale thresholds is about the same with and without exotics provided that they are not too light.  The allowed exotic sectors in our models fall into two categories with one helping to counterbalance distortions of unification from 2-loop running in the MSSM and one making the situation slightly worse.  In both cases, exotic contributions are comparable in size to 2-loop effects.}}

{{The second part of our study concerns the possible impact of exotics on LHC physics.  The exotics come in vector-like pairs with respect to the MSSM and naturally couple to singlet fields that extend into the bulk of the compactification.  It seems plausible that the singlets will be sensitive to supersymmetry-breaking that occurs in the bulk so our exotics naturally play the role of gauge messenger fields.  If we further assume that the gravitino is relatively light, this leads to a variant of gauge mediation in which the messengers do not come in complete GUT multiplets.}}{\footnote{A similar proposal was made recently for exotics that arise in certain Heterotic orbifold models \cite{Anandakrishnan:2011zn}.  There, like here, the exotics did not form complete GUT multiplets.}}.
{{This may seem reminiscent of generalized, not necessarily GUT-like messenger sectors in gauge mediation as in \cite{Martin:1996zb} (see also \cite{Meade:2008wd}), however the crucial difference is that in \cite{Martin:1996zb} the messengers are chosen such they}} {induce  corrections to the running that are universal for all three gauge groups, thereby maintaining 1-loop gauge-coupling unification. 
 This is in contrast to the situation here, where the gauge messengers are non-GUT multiplets which clearly do not contribute as complete GUT multiplets to the running. 
 Nevertheless, as we discussed earlier,  other threshold effects that are mainly due to UV physics, can render these models compatible with the measured weak scale values of the gauge couplings.}}
 {{The messenger scale in our models corresponds to the exotic mass, which must be light enough that $U(1)$-violating operators are not regenerated without significant suppression.  On the other hand, the messenger scale must be heavy enough that the disruption of unification from exotics is minimal.  We typically take the exotic masses to be smaller than the unification scale by a factor of $\sim 100$ so that a variant of high-scale gauge mediated supersymmetry breaking (GMSB) results.}}
 }

{{We proceed in section \ref{sec:phenomenology} to study the phenomenology of high-scale GMSB models with non-GUT messenger sectors that come from the survey of \cite{Dolan:2011iu}.}}
We provide a series of benchmark points whose spectra and phenomenology we study, paying particular interest to prospects and signals at the LHC. We also investigate the global features of the models through parameter scans.  Finally, we investigate how {{the non-universal gaugino masses of these models can help to distinguish them} from minimal GMSB  with complete GUT multiplet messenger fields.

\section{Summary: F-theory GUTs with $U(1)_{PQ}$}
\label{sec:surveysummary}

We now provide a brief summary of the analysis of \cite{Dolan:2011iu} and characterize the class of models that will be analyzed subsequently.  F-theory GUT models can be thought of, from a low energy perspective, as descending from an 8-dimensional supersymmetric Yang-Mills theory associated to the worldvolume of a stack of 7-branes \cite{Donagi:2008ca,Beasley:2008dc,Beasley:2008kw,Donagi:2008kj}.  This worldvolume is compactified on a complex surface $S_{\rm GUT}$ whose volume sets a Kaluza-Klein scale $M_{KK}$, below which we effectively get 4-dimensional physics from dimensional reduction.  The presence of 4 internal dimensions that open up at the scale $M_{KK}$ introduces an elegant way to break the GUT group; we need only turn on a nontrivial field strength for hypercharge, referred to as a `hypercharge flux', on the internal directions \cite{Beasley:2008kw,Donagi:2008kj}.  Off-diagonal `$X$' and `$Y$' gauge bosons acquire a KK scale mass and we identify $M_{KK}$ with the 'unification scale' $M_U\sim M_{KK}$.

Building realistic models requires additional structure to prevent phenomenologically undesirable couplings, such as the tree-level $\mu$ term and dimension 5 proton decay operators
\begin{equation}\mu\int\,d^2\theta\, H_uH_d\qquad\text{ and }\qquad \frac{1}{\Lambda}\int\,d^2\theta\,Q^3L  \,.
\label{dangerousoperators}\end{equation}
Discrete symmetries, such as those recently studied in \cite{Lee:2010gv,Lee:2011dya}, provide one candidate for the extra structure that we need.  An alternative to this has received significant attention in recent years is the introduction of gauged $U(1)$'s \cite{Marsano:2009gv,Blumenhagen:2009yv,Marsano:2009wr,Grimm:2009yu}.  
Technical aspects of engineering $U(1)$'s aside \cite{Hayashi:2010zp,Grimm:2010ez,Marsano:2010ix,Marsano:2011nn,Grimm:2011tb,TimoU1}, one can see already from the 8-dimensional worldvolume theory{\footnote{These conclusions require a very slight knowledge of physics from the bulk, namely the way that bulk axions couple to the brane worldvolume and the mechanism for canceling gauge anomalies involving $U(1)$'s that are lifted by bulk flux through the Stuckelberg mechanism}} that combining $U(1)$'s and `hypercharge flux' places severe restrictions on the theory.  This can be summarized by the Dudas-Palti relations \cite{Dudas:2010zb,Marsano:2010sq}, which constrain the `hypercharge flux' that threads distinguished curves $\Sigma_R$ on $S_{\rm GUT}$ associated to the appearance of chiral multiplets in the representation $R$
\be
\sum_{\mathbf{10}\text{ matter curves, }i}q_i\int_{\Sigma_{\mathbf{10}}^{(i)}}F_Y = \sum_{\mathbf{\overline{5}}\text{ matter curves, }a}q_a\int_{\Sigma_{\mathbf{\overline{5}}}^{(a)}} F_Y \,.
\label{DP}
\ee
Here, $q_i/q_a$ denotes the common $U(1)$ charge carried by chiral multiplets in the $\mathbf{10}/\mathbf{\overline{5}}$ representation associated with the curve $\Sigma_{\mathbf{10}}^{(i)}/\Sigma_{\mathbf{\overline{5}}}^{(a)}$.There are two simple, but consequential, implications of these relations:

\begin{itemize}
\item If we require precisely the spectrum of the MSSM with no additional exotic particles, there is a unique (up to normalization) flavor-blind $U(1)$ consistent with \eqref{DP} that preserves the MSSM superpotential.  This is the famous $U(1)_{\chi}$ that descends from $SO(10)$ and is given by a linear combination of $U(1)_{B-L}$ and $U(1)_Y$
\begin{equation}\begin{array}{c|c|c|c|c}
 & \mathbf{10}_M & \mathbf{\overline{5}}_M & H_u & H_d \\ \hline
U(1)_{\chi} & 1 & -3 & -2 & 2
\end{array}\end{equation}
This symmetry does not protect us from either dimension 5 proton decay or the possible generation of a $\mu$ term.

\item If we require a $U(1)_{PQ}$ symmetry, that is a symmetry for which
\be
q_{PQ} (H_u) + q_{PQ} (H_d) \not= 0 \,,
\ee
then \eqref{DP} implies the presence of quasi-chiral{\footnote{Following very clever terminology we learned from P.~Langacker, we use quasi-chiral to denote a collection of fields that is chiral with respect to $U(1)_{PQ}$ but vector-like with respect to the MSSM gauge group.}} exotic fields that do not come in complete $SU(5)$ multiplets{\footnote{The presence of exotics like this was already noted in \cite{Marsano:2009gv,Marsano:2009wr}}}.  A $U(1)_{PQ}$ symmetry has the advantage of helping with the $\mu$ and dimension 5 proton decay problems but now the exotics must be dealt with.
\end{itemize}

In this paper, we set out to study the physics of $U(1)_{PQ}$ models and the consequences of exotics that are forced on us by \eqref{DP}.  Since the exotics are vector-like with respect to the MSSM, they can be removed from the spectrum via a superpotential coupling
\be\label{Xffb}
W \supset X f_{ex} \bar{f}_{ex} 
\ee
provided the field $X$, which is an MSSM singlet that carries $PQ$ charge, acquires a nontrivial expectation value.  We have to be careful about how large this expectation value gets because $\langle X\rangle$ sets the scale of PQ-breaking that we should expect in the low energy effective action.  If $\langle X\rangle$ becomes too large, the dangerous operators \eqref{dangerousoperators} will not be sufficiently suppressed.

A detailed look at the structure of possible exotic sectors can be found in \cite{Dolan:2011iu}; we focus here on exotic sectors that are consistent with \eqref{DP} and satisfy the further condition that all exotics can be lifted by the vev of a single field $X$, thereby acquiring roughly equivalent masses.  In this case, we can use the relation \eqref{DP} to show that the exotic spectrum can be parametrized by four integers $M$, $N$, $K$, and $L$ as

\begin{equation}\begin{array}{c|c|c}\label{KLMNLabels}
SU(5)\text{ origin} & \text{Exotic Multiplet} & \text{Degeneracy}  \\ \hline
& (\mathbf{1},\mathbf{1})_{+1}\oplus (\mathbf{1},\mathbf{1})_{-1} & M+N \\
\mathbf{10}\oplus\mathbf{\overline{10}} & (\mathbf{3},\mathbf{2})_{+1/6}\oplus (\mathbf{\overline{3}},\mathbf{2})_{-1/6} & M \\
& (\mathbf{\overline{3}},\mathbf{1})_{-2/3}\oplus (\mathbf{3},\mathbf{1})_{+2/3} & M-N\\ \hline
\mathbf{\overline{5}}\oplus \mathbf{5} & (\mathbf{\overline{3}},\mathbf{1})_{+1/3} \oplus (\mathbf{3},\mathbf{1})_{-1/3} & K \\
& (\mathbf{1},\mathbf{2})_{-1/2}\oplus (\mathbf{1},\mathbf{2})_{+1/2} & K-L \\
\end{array}\end{equation}
where consistency of the parametrization requires
\begin{equation}
M\ge |N|\,,\qquad K\ge \text{min}(0,L)\,.
\label{exoticparam}\end{equation}
Further, \eqref{DP} tells us the power of $X$ that appears in nonrenormalizable operators that can generate \eqref{dangerousoperators}. 
More specifically, the operators
\begin{equation}\frac{1}{\Lambda}\int\,d^2\theta\,\left(\frac{X}{\Lambda}\right)^{\Delta}Q^3L\qquad\hbox{and }\qquad  \Lambda\int\,d^2\theta\,\left(\frac{X}{\Lambda}\right)^{-\Delta}H_uH_d
\label{regrownoperators}\end{equation}
are always gauge invariant where
\begin{equation}\Delta\equiv N-L\,.
\end{equation}
A bosonic expectation value for $X$ can grow the proton decay operator if $N-L>0$ while it will grow a $\mu$ parameter if $N-L<0$.  Because the $\mu$ problem requires more severe suppression, $N-L>0$ is in some sense preferred{\footnote{Note that if $X$ picks up an $F$-component expectation value, the operator from \eqref{dangerousoperators} that isn't generated by the bosonic vev $\langle X\rangle$ can be grown from K\"ahler potential corrections involving $X^{\dag}$.  In the case of $\mu$, this is reminiscent of the Giudice-Masiero mechanism \cite{Giudice:1988yz} and represents a nice solution to the $\mu$ problem \cite{Ibe:2007km} that has been suggested in the context of F-theory GUT models before \cite{Marsano:2008jq,Heckman:2008qt}.}}.

So far we have only used the relations \eqref{DP}.  If we ask which choices of $(M,N,K,L)$ can actually be realized in explicit constructions, then considerations based on the 8-dimensional worldvolume gauge theory alone already limit us to a small set of possibilities \cite{Dolan:2011iu}
\begin{equation}\begin{array}{c|cc|c}
\text{Models} & \text{Exotic Spectra} & & \text{Dim 5} \\ \hline
I  & N-L=1 & & XQ^3L/\Lambda^2 \\
II & N-L=2 & K\ge M & X^2Q^3 L / \Lambda^3 \\
III & L=2 & M=N=0 & X^{\dag\,2}Q^3L /\Lambda^4 \\
IV & N-L=1 & K-L=M & XQ^3L/\Lambda^2
\end{array}\end{equation}
The last column denotes the operator responsible for generating dimension 5 proton decay whose power of $X$ is $\Delta=N-L$ as we said above.
It is curious that 
\begin{equation}\Delta = N-L=-2,1,2\label{DeltaVals}\end{equation}
all appear but $N-L=-1$ does not.
{{For the phenomenological studies,  we will consider the possibility that the singlet $X$ from which the exotics derive their mass also obtains an $F$-component expectation value.  Supersymmetry-breaking will be communicated to the visible sector through loops of exotic fields. If we assume this is the dominant method of communication then we arrive at a scenario in which the exotics are generalized messengers of gauge mediation.  Indeed, models with PQ symmetries  -- albeit with messengers in GUT multiplets -- were studied in local F-theory models in the early days \cite{Marsano:2008jq, Heckman:2008qt} following \cite{Ibe:2007km}. 
}}

\section{Unification}
\label{app:unification}

Since our models add new fields to the MSSM that do not come in complete $SU(5)$ multiplets, it is necessary to consider the impact of these fields on unification.  In this section, we review the unification story in F-theory GUT models in general and study the impact of exotic fields that appear in our $U(1)_{PQ}$ models.

\subsection{Generalities and Overview}
\label{subsec:unigeneralities}

The issue of unification in F-theory GUTs has received some attention in recent years \cite{Donagi:2008kj,Blumenhagen:2008aw,Leontaris:2011pu,Leontaris:2011tw,Callaghan:2011jj}.  We begin by recalling a few results of this previous work and other general features of precision unification studies that will enter our analysis.  A universal feature of F-theory GUTs that utilize hypercharge flux to break the GUT group \cite{Beasley:2008kw,Donagi:2008kj} is the presence of flux-induced splittings of MSSM gauge couplings at the KK scale \cite{Donagi:2008kj,Blumenhagen:2008aw}.  The treatment of \cite{Blumenhagen:2008aw} looked at these splittings in type IIB orientifold GUT models by studying the D7-brane worldvolume coupling
\begin{equation}\int_{\mathbb{R}^{3,1}\times S_{\rm GUT}}\text{tr}_f F^4\,,
\label{7branecoupling}\end{equation}
where the trace is taken in the fundamental representation.  Through \eqref{7branecoupling} one can see that a nontrivial flux $F_Y$ on $S_{\rm GUT}$ in the hypercharge direction will generate non-universal contributions to the effective 4-dimensional gauge couplings upon performing the integral over $S_{\rm GUT}$.  These splittings carry a special structure in that the high scale gauge couplings satisfy the Blumenhagen relation \cite{Blumenhagen:2008aw}
\begin{equation}\alpha_1(M_{KK}) - \frac{3}{5}\alpha_2(M_{KK}) - \frac{2}{5}\alpha_3(M_{KK})=0\,.
\label{blumenhagenrelation}\end{equation}
As pointed out in \cite{Blumenhagen:2008aw}, another way to induce splittings with this structure is to introduce an extra vector-like pair of triplets.  This is quite nice from the perspective of the models that we study because a vector-like pair of triplets corresponds to the simple parameter choice $(K,L,M,N)=(1,1,0,0)$.  In fact, the situation seems even better than this; regardless of the choice of $(K,L,M,N)$, all of the exotic sectors that we consider lead to 1-loop $\beta$ function shifts that satisfy
\begin{equation}\delta\beta_1 - \frac{3}{5}\delta\beta_2 - \frac{2}{5}\delta\beta_3  =0\,.
\end{equation}
As emphasized in \cite{Marsano:2009gv,Marsano:2009wr}, this makes the idea of balancing the splitting due to hypercharge flux with splittings from exotics seem quite plausible.

Unfortunately, the detailed unification story is not so elegant.  Firstly, the extent to which the relation \eqref{blumenhagenrelation} describes the structure of splittings in F-theory models is a bit subtle \eqref{blumenhagenrelation}\footnote{On this point, we are very grateful to M.~Wijnholt for valuable discussions and for sharing with us a draft revision of \cite{Donagi:2008kj} where this issue is addressed.}.  As observed by Donagi and Wijnholt \cite{Donagi:2008kj}, one finds that the 7-brane worldvolume theory exhibits a coupling similar to \eqref{7branecoupling} but with the fundamental trace replaced by a trace over the adjoint reprsentation \cite{Donagi:2008kj}
\begin{equation}\int_{\mathbb{R}^{3,1}\times S_{\rm GUT}}\text{tr}_{adj}F^4=\int_{\mathbb{R}^{3,1}\times S_{\rm GUT}} \left(10\text{tr}_fF^4 + 6\left[\text{tr}_f(F^2)\right]\right) \,.\label{logsplittings}\end{equation}
To be concrete, this coupling arises as the coefficient of a `universal' logarithmic divergence of the 7-brane worldvolume theory \cite{Donagi:2008kj} that must be cut off at some bulk `winding' scale $M_{\rm Bulk}$ \cite{Conlon:2009xf,Conlon:2009qa}.  The structure of the splittings that result is equivalent to what one would obtain by integrating out a vector-like pair of lepto-quarks in the $(\mathbf{3},\mathbf{2})_{-5/6}\oplus(\mathbf{\overline{3}},\mathbf{2})_{+5/6}$ representations that descend from the $SU(5)$ adjoint.  As we review in the discussion of high scale thresholds in section \ref{subsec:highthresh}, a relation of the type \eqref{blumenhagenrelation} does emerge in a certain sense but this requires the introduction of a new fictitious scale $M_{fict}$ that is neither $M_U$ nor the bulk scale $M_{\rm Bulk}$.  
This scale is helpful to understand the relative sizes of splittings from exotics and those induced by \eqref{logsplittings} and can be used to show that a precise cancellation of these as suggested by \cite{Marsano:2009gv,Marsano:2009wr} is actually not possible.  One of our primary interests will be to ensure that physical scales $M_U$ and $M_{\rm Bulk}$ satisfy certain consistency conditions, though, so we will mostly choose not to discuss the unphysical scale $M_{fict}$.

It is not a disaster that exotics and logarithmic KK corrections cannot cancel one another because there are a number of additional corrections that we must include.  Most obvious are standard corrections from the MSSM including low scale thresholds and 2-loop running.  In addition, however, there are finite threshold effects at the high scale that arise from explicit
computations of Ray-Singer torsions associated to the hypercharge bundle on $S_{\rm GUT}$ and the various matter curves $\Sigma$ \cite{Donagi:2008kj}.  We expect these corrections to be highly geometry and bundle dependent so, from a 4-dimensional point of view they represent corrections that are sensitive to unknown details of UV physics.  Our approach is to introduce model parameters to encapsulate these quantities.  In the end, our analysis will amount to introducing bounds on these parameters that are necessary to accomplish the following goals:
\begin{itemize}
\item Attain consistency with the observed low scale gauge couplings
\item Ensure that the 'winding' scale $M_{\rm Bulk}$ is larger than the true 'unification scale' $M_U$, which receives corrections relative to the 1-loop GUT scale $M_U^{(0)}\sim 2\times 10^{16}$ GeV and is identified with the KK scale of the 7-brane theory $M_U\sim M_{KK}$
\item Keep the exotic mass smaller than $10^{-2}M_U$
\end{itemize}
The last condition is necessary because the exotic mass (which is roughly $\langle X\rangle$ from \eqref{Xffb}) sets the scale of $PQ$ violation.  If it becomes too large, we risk growing the operators \eqref{dangerousoperators} from nonrenormalizable couplings of the form \eqref{regrownoperators}.  As it is, we will probably need some further tuning even when $M_{\text{Exotic}}\sim 10^{-2}$ but it isn't so bad when $\Delta = N-L=-2$ so that $\mu$ is not generated and dimension 5 proton decay gets quadratic suppression.  Let us also comment that $M_{\text{Exotic}}<10^{-2}M_U$ is also required if we want to effectively separate the physics induced by exotic fields through loops from physics generated by the KK tower.  The gauge mediated scenarios of section \ref{sec:phenomenology} rely on an assumption of this type.

The bounds that we obtain will depend on the flux choices $(K,L,M,N)$ and sparticle masses.  We will display these bounds for two sample flux choices under the assumption that sparticle masses are generated by a gauge mediation scenario in which the exotics play the role of gauge messenger fields.  Reasonable scale separations between $M_{\text{Exotic}}$ and $M_U$ can be achieved with finite threshold parameters that are of ${\cal{O}}(1)$.  
When $N-L<0$, exotics actually decrease the requisite size high scale unknowns relative to what one would need to deal with 2-loop distortions alone while for $N-L>0$ the requisite size is slightly larger but not much so.  In both cases, $|N-L|\le 2$ requires threshold parameters that are ${\cal{O}}(1)$, which we believe is quite plausible.

In the rest of this section we review 1-loop running in the MSSM to set conventions before reviewing how we describe corrections to the 1-loop story and discussing contributions from high scale thresholds and exotics.  The standard MSSM corrections from low scale thresholds and 2-loop running are reviewed in Appendix \ref{app:RG}.


\subsection{1-loop MSSM Running}

We start by reviewing the basics of 1-loop gauge coupling unification in the MSSM.  This will allow us an opportunity to set notation as well as normalization and sign conventions.  At 1-loop in the MSSM, the gauge couplings run as
\begin{equation}\frac{1}{\alpha_i(m_Z)} = \frac{1}{\alpha_i(M_U)} + \frac{b_i}{2\pi}\ln\left(\frac{M}{m_Z}\right) \,,
\end{equation}
where the $\beta$ function coefficients are
\begin{equation}\begin{split}
\begin{pmatrix} b_1 \\ b_2 \\ b_3 \end{pmatrix} &= \begin{pmatrix} 0 \\ -6 \\ -9\end{pmatrix} + N_{gen}\begin{pmatrix} 2 \\ 2 \\ 2\end{pmatrix} + N_{Higgs}\begin{pmatrix}3/10 \\ 1/2 \\ 0\end{pmatrix}\\
&=\begin{pmatrix} 33/5 \\ 1 \\ -3\end{pmatrix} \,.
\end{split}\label{MSSMbeta}\end{equation}
Here $N_{gen}$ is the number of generations, $N_{Higgs}$ is the number of Higgs doublets, and in the last line we have set $N_{gen}=3$ and $N_{Higgs}=2$.  
As usual, we proceed by supposing that a unified coupling $\alpha_{U}$ is specified at a fixed scale $M_U$.  Given this, we will see how well we can fit "data" by adjusting the scale $M_U$ and the unified coupling constant, $\alpha_{U}$, at that scale.  The conventional set of data to use are not the independent gauge couplings $\alpha_i$ but rather the observable quantities $\sin^2\theta_W(m_Z)$, $\alpha_{em}(m_Z)$, and $\alpha_3(m_Z)$.  We quickly recall that, in our notation and conventions
\begin{equation}\sin^2\theta_W = \frac{3\alpha_1}{3\alpha_1+5\alpha_2}\qquad \alpha_{em} = \frac{3\alpha_1\alpha_2}{3\alpha_1+5\alpha_2}\,.\end{equation}
The measured values to which we compare are
\begin{equation}\begin{split}\sin^2\theta_W(m_Z)&\sim 0.2312\pm 0.0002 \\
\alpha_{em}^{-1}(m_Z)&\sim 127.906\pm 0.019 \\
\label{thetawalphaem}
\end{split}\end{equation}
and the strong coupling constant evaluated at $M_Z$, $\alpha_3(m_Z)$. The 2009 World Average for this figure is given by Bethke~\cite{Bethke:2009jm} as \begin{equation}\alpha_3(m_Z) \sim 0.1184\pm 0.0007.\label{alphastrong}\end{equation}

As the 1-loop MSSM RGE's have only two free dimensionless parameters, $\alpha_{U}$ and $M_{U}/m_Z$, the GUT framework can predict one of the 3 parameters of \eqref{thetawalphaem} and \eqref{alphastrong}.  It is conventional to choose $\alpha_U$ and $M_U/m_Z$ to fit $\alpha_{em}$ and $\sin^2\theta_W$ and use these values to `predict' $\alpha_3$.  We will use the notation $M_U^{(0)}$ and $\alpha_U^{(0)}$ for the 1-loop values of $M_U$ and $\alpha_U$ that are used to fit the observed $\sin^2\theta_W(m_Z)$ and $\alpha_{em}^{-1}(m_Z)$.   We have
\begin{equation}\begin{split}\frac{1}{2\pi}\ln \frac{M_{U}^{(0)}}{m_Z} &= \frac{1}{b_1-b_2}\left(\alpha_1^{-1}(m_Z)-\alpha_2^{-1}(m_Z)\right) \\
&= \alpha_{em}^{-1}(m_Z)\times\left[\frac{3-8\sin^2\theta_W(m_Z)}{5(b_1-b_2)}\right]\\
\alpha_{U}^{(0),\,-1} &= \frac{b_1\alpha_2^{-1}(m_Z) - b_2\alpha_1^{-1}(m_Z)}{b_1-b_2} \\
&= \alpha_{em}^{-1}(m_Z)\times\left[ \frac{5b_1\sin^2\theta_W(m_Z) - 3b_2\cos^2\theta_W(m_Z)}{5(b_1-b_2)}\right] \,.
\end{split}\label{alphaMU1loop}\end{equation}
Plugging this into the RG running for $\alpha_3$ leads to
\begin{equation}\begin{split}\alpha_3^{-1}(m_Z) &= \alpha_{U}^{(0),\,-1} + \frac{b_3}{2\pi}\ln\left(\frac{M_U^{(0)}}{m_Z}\right) \\
&= \alpha_{em}^{-1}(m_Z)\times\left[\frac{5b_1\sin^2\theta_W(m_Z) - 3b_2\cos^2\theta_W(m_Z) + b_3\left(3-8\sin^2\theta_W(m_Z)\right)}{5(b_1-b_2)}\right] \,.
\end{split}\end{equation}
Inserting the measured values for $\sin^2\theta_W$ and $\alpha_{em}$ and propagating the experimental errors through the calculation yields
\begin{equation}\alpha_3(m_Z) \sim 0.1169 \pm 0.0002 \,,\end{equation}
which disagrees with the measured value by only about 1.5\%.  As we know very well, unification works great in the MSSM at 1-loop level.


\subsection{{{General}} Parametrization of Corrections}
\label{sec:generalities}

We now outline our general approach to parametrizing corrections which follows that of \cite{Alciati:2005ur}.  As already noted in the introduction to this section, we will incorporate all corrections beyond 1-loop running by parameters $\delta_i$ that enter into the gauge couplings as
\begin{equation}\alpha_i^{-1}(m_Z) = \alpha_U^{-1} + \frac{b_i}{2\pi}\ln\left(\frac{M_U}{m_Z}\right) + \delta_i\,.
\end{equation}
Redoing the above analysis with these additional corrections, we find that
\begin{equation}\begin{split}
\frac{1}{2\pi}\ln\frac{M_{U}}{m_Z} &= \frac{1}{b_1-b_2}\times\left[\alpha_1^{-1}(m_Z) - \alpha_2^{-1}(m_Z)-\delta_1+\delta_2\right] \\
&= \frac{1}{2\pi}\ln\frac{M_U^{(0)}}{m_Z} + \delta\left(\frac{1}{2\pi}\ln\frac{M_U}{m_Z}\right) \\
\alpha_{U}^{-1} &= \frac{1}{b_1-b_2}\times\left[ b_1\left(\alpha_2^{-1}(m_Z)-\delta_2\right) -b_2\left(\alpha_1^{-1}(m_Z) - \delta_1\right)\right] \\
&= \alpha_U^{(0)\,-1} + \delta\left(\alpha_U^{-1}\right) \,,
\end{split}\label{LLaGUT}\end{equation}
where we recall that $M_U^{(0)}$ and $\alpha_U^{(0)}$ denote the 1-loop values \eqref{alphaMU1loop} and we have defined the shifts in these quantities as
\begin{equation}\begin{split}\delta\left(\frac{1}{2\pi}\ln\frac{M_U}{m_Z}\right) &= \frac{\delta_2-\delta_1}{b_1-b_2} \\
\delta\left(\alpha_U^{-1}\right) &= \frac{b_2\delta_1-b_1\delta_2}{b_1-b_2} \,.
\end{split}\end{equation}
It may happen that the $\delta_i$ depend on $M_U$ and $\alpha_U$.  In that case, we can get the leading corrections by using the 1-loop values, $M_U^{(0)}$ and $\alpha_U^{(0)}$.

With the above results, the ``predicted" value for $\alpha_3(m_Z)^{-1}$ is
\begin{equation}\begin{split}\alpha_3(m_Z)^{-1} &= \frac{1}{5(b_1-b_2)\alpha_{em}(m_Z)}\times\left[3(b_3-b_2)\cos^2\theta_W(m_Z) + 5(b_1-b_3)\sin^2\theta_W(m_Z)\right] + \Delta\\
&= \alpha_3(m_Z)^{-1,(1-loop)} + \Delta\end{split}\end{equation}
with
\begin{equation}\Delta \equiv \left[\frac{(b_2-b_3)\delta_1 + (b_3-b_1)\delta_2}{b_1-b_2}+\delta_3\right] \,.
\end{equation}
Inverting the expression for $\alpha_3(m_Z)$ at 1-loop we find
\begin{equation}\alpha_3(m_Z) = \alpha_3^{(1-loop)}(m_Z)\times\left(1-\alpha_3^{(1-loop)}(m_Z)\Delta+\ldots\right) \,. \label{eq:predMZ}\end{equation}
Because $\alpha_3^{(1-loop)}(m_Z)$ is about 1.5\% too small, we would like the correction term to be negative and around $1.5\times 10^{-2}$.  To nail the experimental result right on the head we would actually need
\begin{equation}\Delta^{(\text{ideal})}(m_Z) \sim -0.11 \pm 0.05 \,,\label{Deltaideal}\end{equation}
where we have included the errors on $\alpha_3^{(1-loop)}(m_Z)$ and $\alpha_3^{(meas)}(m_Z)$.
Note that for the MSSM $\beta$ functions we get
\begin{equation}\Delta = \frac{1}{7}\left[5\delta_1 - 12\delta_2 + 7\delta_3\right]\,.
\end{equation}

We will apply these expressions to determine the relatively simple corrections due to 2-loop MSSM running, exotics and low-scale thresholds in appendix A, and in the next subsections to determine the corrections due to high-scale thresholds and the non-GUT exotics.  


\subsection{High Scale Thresholds}
\label{subsec:highthresh}

Let us now consider high scale corrections 
that are generated by integrating out KK modes of the 8-dimensional brane worldvolume theory.  For this, we use the results of Donagi and Wijnholt\cite{Donagi:2008kj}{\footnote{Again, we are very grateful to M.~Wijnholt for sharing a draft revision of \cite{Donagi:2008kj} with us that contains these results.}}.

\subsubsection{High Scale Threshold Corrections in F-theory GUTs}

As F-theory models cease to be 4-dimensional at the KK scale $M_{KK}=M_U$, degrees of freedom with masses $M>M_{U}$ must be integrated out.  These modify the gauge kinetic functions and, in turn, the effective 4-dimensional gauge couplings specified at $M_U$.  One might expect that these corrections only renormalize the unified coupling constant $\alpha_U$ but splittings between the different MSSM couplings are induced when an internal hypercharge flux is used to break $SU(5)$ because that flux has a nontrivial impact on the 7-brane KK spectrum.  There is no sense in which these corrections are small either since one encounters logarithmic divergences when integrating out the KK modes that do not cancel.  This reflects the fact that the 7-brane theory on its own is not consistent without incorporating bulk physics to cancel `local tadpoles' \cite{Conlon:2009xf,Conlon:2009qa}.  Logarithmic divergences from KK modes, as well as from MSSM fields, are capped not at $M_{U}$ but rather at some bulk scale $M_{\rm Bulk}$ that the authors of \cite{Conlon:2009xf,Conlon:2009qa} refer to as the `winding scale'.

Because the divergences are logarithmic, they can be written in a way that effectively mimics a `fake' 4-dimensional RG flow from above the KK scale as in \cite{Conlon:2009xf,Conlon:2009qa}.  We see this explicitly by noting that the effect of these divergences is to replace the unified coupling constant $\alpha_U$ at the KK scale $M_U$ by
\begin{equation}\alpha_U^{-1}\rightarrow \alpha_i(M_U)^{-1}+\frac{b_i^{(KK)}}{2\pi}\ln\left(\frac{M_{\rm Bulk}}{M_U}\right)+\text{finite}\label{highscalegeneral}\,.\end{equation}
We also indicate here the presence of finite corrections from high scale effects whose computation must be performed on a model-by-model basis.  When necessary, we will introduce new parameters to encapsulate our ignorance of these terms.

The corrections from KK modes are computed by Ray-Singer torsions associated to the hypercharge bundle on the surface $S_{\rm GUT}$ and the distinguished curves $\Sigma$ associated to matter multiplets.  These, in turn, have a universal scaling behavior that allowed Donagi and Wijnholt to determine the coefficients $b_i^{(KK)}$ of the logarithmic divergences in \eqref{highscalegeneral} \cite{Donagi:2008kj}
\begin{equation}\begin{split}
b_i^{(KK)} &= \sum_{R\in\text{Adj}} (2\delta b_R) \int_{S_{\rm GUT}}\left[-\frac{1}{2}\text{ch}_2(V_R) - \frac{1}{6}c_1(V_R)c_1(T_{S_{\rm GUT}})-\frac{1}{24}c_2(T_{S_{\rm GUT}})\right] \\
&\quad + \sum_{\text{Matter curves, }\Sigma}\sum_{R\in R_{\Sigma}} (2\delta b_R) \left[-\frac{1}{12}\chi(\Sigma)\right]
\end{split}\label{DWresult}\,.\end{equation}
Here the sum in the first line is over all MSSM representations that sit in the adjoint of $SU(5)$ while the sum in the second line is over all MSSM representations that sit in a representation $R_{\Sigma}$ of $SU(5)$ associated to a particular matter curve, $\Sigma$.  The coefficients $\delta b_R$ correspond to the shift in the MSSM $\beta$ function coefficients, in our sign and normalization conventions, from a chiral superfield in the representation $R$.  Finally, $V_R$ is the gauge bundle to which the representation $R$ couples.  Depending on the representation $R$, we can always take $V_R$ to be a suitable power of the hypercharge bundle.  

The only non-universality in these shifts comes from the dependence on $V_R$, which allows us to immediately conclude that no logarithmic non-universalities are generated by KK modes on matter curves $\Sigma$.  In general we still expect finite contributions from threshold corrections associated to KK modes on matter curves, though even these may vanish in special circumstances \cite{Leontaris:2011tw}.  Returning to logarithmic contributions, the only non-universal ones come from KK modes that propagate on the full 7-brane worldvolume $S_{\rm GUT}$ that descend from the $SU(5)$ adjoint.  Of these, the only modes that couple to the hypercharge bundle are the $(\mathbf{3},\mathbf{2})_{-5/6}$'s and their conjugates in the first line of \eqref{DWresult} for which $V_R=V_Y$, the hypercharge bundle. 

To evaluate the $V_Y$-dependent contribution from these we need two further facts.  The first is that $\int_{S_{\rm GUT}}c_1(V_Y)\cdot c_1(T_{S_{\rm GUT}})=0$ which is a consequence of ensuring that $V_Y$ doesn't lift the hypercharge gauge boson that can be understood, for instance, from anomaly considerations \cite{Donagi:2008kj,Marsano:2010sq,Dolan:2011iu}.  The second is that $\int_{S_{\rm GUT}}\text{ch}_2(V_Y)=-1$ which is a necessary condition for ensuring that there are no massless modes in the $(\mathbf{3},\mathbf{2})_{-5/6}$ representation or its conjugate.  Taken together, this means that the effective $\beta$ function shifts from KK mode high scale thresholds are simply those that we would obtain from a single vector-like pair of chiral superfields in the representation $(\mathbf{3},\mathbf{2})_{-5/6}\oplus (\mathbf{\overline{3}},\mathbf{2})_{+5/6}$
\begin{equation}b_i^{(KK)} = b_i^{(\mathbf{3},\mathbf{2})_{-5/6}\oplus\text{cc}} = \begin{pmatrix}5 \\ 3 \\ 2\end{pmatrix}\label{betaKK}\,.\end{equation}
As Donagi and Wijnholt point out \cite{Donagi:2008kj}, this can also be understood from the fact that the logarithmic divergence of the 7-brane worldvolume theory is proportional to
\begin{equation}\int_{\mathbb{R}^{3,1}\times S_{\rm GUT}}\text{tr}_{Adj}F^4= \int_{\mathbb{R}^{3,1}\times S_{\rm GUT}}\left(10\text{tr}_fF^4+6\left[\text{tr}_fF^2\right]^2\right)\,.\end{equation}
A study of this term in the presence of a nontrivial hypercharge flux $F_Y$ along the lines of \cite{Blumenhagen:2008aw} leads to non-universal $\beta$ function shifts proportional to \eqref{betaKK}.

We are almost done with high scale thresholds but one correction remains to be included.  Divergences from KK modes are not the only ones that are capped off at the scale $M_{\rm Bulk}$.  The contributions from massless fields running in loops also generate divergences and these too must be capped off at the bulk scale $M_{\rm Bulk}$ \cite{Conlon:2009xf,Conlon:2009qa}.  The net high scale corrections that we obtain, then, are given by
\begin{equation}\delta_i^{(\text{High Thresh})} = \frac{b_i^{(\mathbf{3},\mathbf{2})_{-5/6}\oplus\text{cc}}}{2\pi}\ln\left(\frac{M_{\rm Bulk}}{M_U^{(0)}}\right) + \frac{b_i^{(MSSM)}}{2\pi}\ln\left(\frac{M_{\rm Bulk}}{M_U^{(0)}}\right)\,.\end{equation}
We can now compute the contribution of high scale thresholds to $\Delta$ as well as the shift in $M_U$ relative to the 1-loop value $M_U^{(0)}$.  We also introduce parameters $\Delta^{(finite)}$ and $\delta_{M_U}^{(finite)}$ to represent finite contributions to these quantities from high scale threshold corrections.  We do not discuss $\alpha_U^{-1}$ because this will also be significantly renormalized by high scale thresholds and, in the end, its precise value will not be of much importance to us.  For $\Delta$ and the shift in $M_U$ we obtain
\begin{equation}\begin{split}\Delta^{(\text{High Thresh})} &= \frac{3}{14\pi}\ln\frac{M_{\rm Bulk}}{M_U^{(0)}} +\Delta^{(finite)}\\
\delta\left(\frac{1}{2\pi}\ln\frac{M_U}{m_Z}\right)^{(\text{High Thresh})} &= -\frac{19}{28\pi}\ln\frac{M_{\rm Bulk}}{M_U^{(0)}}+ \delta_{M_U}^{(finite)}
\end{split}\label{highthreshcorrs}\,.\end{equation}
We do not know much about the finite parameters $\Delta^{(finite)}$ or $\delta_{M_U}^{(finite)}$ but \cite{Donagi:2008kj} studied some toy models and evaluated explicit Ray-Singer torsions in the simple case that $S_{\rm GUT}=\mathbb{P}^1\times \mathbb{P}^1$.  In the example there, $\Delta^{(finite)}=\frac{9}{8\pi}\sim 0.36$.  To the extent that we can draw any conclusions from just one data point, it seems reasonable to expect that these finite parameters are neither small enough to be negligible nor large enough that we can be happy with parameter ranges in which they differ much from being of order 1 or so.


\subsubsection{Aside on \eqref{blumenhagenrelation}}
\label{subsubsec:blumenhagenrelation}

Finally, let us return to the Blumenhagen relation \eqref{blumenhagenrelation} that arose from the IIB analysis of \cite{Blumenhagen:2008aw}.  As we mentioned in section \ref{subsec:unigeneralities}, the logarithmic corrections in \eqref{betaKK} do not satisfy this relation but there is a sense in which it naturally emerges.  
This is well known \cite{Donagi:2008kj,Leontaris:2011pu,Leontaris:2011tw} and depends on the simple fact that the $\beta$ function shifts due to leptoquarks $(\mathbf{3},\mathbf{2})_{-5/6}\oplus\text{cc}$ satisfy 
\begin{equation}b_i^{(\mathbf{3},\mathbf{2})_{-5/6}\oplus\text{cc}} = \frac{1}{3}b_i^{(MSSM\text{ gauge})} + 5 \qquad\qquad\qquad b_i^{(MSSM\text{ gauge})}=\begin{pmatrix} 0 \\ -6 \\ -9\end{pmatrix}\label{KKandgauge}\,,\end{equation}
where $b_i^{(MSSM\text{ gauge})}$ are the $\beta$ function shifts induced by MSSM vectors running in the loop.  Because of \eqref{KKandgauge}, the logarithmic contribution of \eqref{highscalegeneral} naturally combines with the ordinary 1-loop MSSM running of vectors in a way that mirrors a simple shift in the UV cutoff scale for that running
\begin{equation}\begin{split}\alpha(m_Z)^{-1} &= \alpha_U^{-1}+\frac{b_i^{(\text{Matter})}}{2\pi}\ln\frac{M_{\rm Bulk}}{M_U}+\frac{b_i^{(MSSM \text{ gauge})}}{2\pi}\ln\frac{M_{fict}}{M_U} + \text{universal} + \ldots\\
&= \alpha_U^{-1}+\frac{b_i^{(\text{Matter})}}{2\pi}\ln\frac{M_{fict}}{M_U}+\frac{b_i^{(MSSM\text{ gauge})}}{2\pi}\ln\frac{M_{fict}}{M_U}+\left[-\frac{b_i^{(\text{Matter})}}{2\pi}\ln\frac{M_{fict}}{M_{\rm Bulk}}+\text{universal}+\ldots\right] \end{split}\label{introduceMfict}\,.\end{equation}
Here the $\ldots$ denotes standard corrections including the 2-loop and low-scale thresholds that are reviewed in Appendix \ref{app:RG}.  The fictitious scale $M_{fict}$ is related to the physical cutoff scale $M_{\rm Bulk}$ by
\begin{equation}M_{fict} = \left(\frac{M_{\rm Bulk}}{M_U}\right)^{1/3}M_{\rm Bulk}\label{Mfict}\,.\end{equation}
From the second line of \eqref{introduceMfict}, we see that the logarithmic KK threshold corrections can be thought of as replacing the bulk scale $M_{\rm Bulk}$ with the fictional scale $M_{fict}$ at the cost of replacing the correction proportional to $b_i^{(\mathbf{3},\mathbf{2})_{-5/6}\oplus\text{cc}}$ with one proportional to $b_i^{(\text{Matter})}$.  The nice thing about $b_i^{(\text{Matter})}$ is that the only nonuniversal contribution comes from the pair of Higgs doublets so it is manifestly obvious that a pair of Higgs doublets or triplets at the right mass can combine with this to give something universal.  More generally, the $b_i^{(\text{Matter})}$ satisfy the Blumenhagen relation \eqref{blumenhagenrelation}.  As we will see shortly, contributions from exotics in our models also satisfy this relation so it naively seems possible, in principle, that they can cancel the explicitly written correction term in \eqref{introduceMfict}, up to universal shifts.

We would have the potential for a very elegant scenario if there were no further corrections.
Even if we assume that the low scale thresholds are negligible, though, we still have finite high scale threshold corrections as well as 2-loop contributions which are not negligible at all.  Ultimately, we must include all of these contributions and, once we start doing that, the motivation for introducing the fictitious scale $M_{fict}$ starts to decrease.  The exotics of our models and the logarithmic KK corrections do not represent two contributions that should cancel one another but rather two pieces to a larger and significantly more muddled puzzle.  In the end, we will have to study how 2-loop issues in the MSSM can be ameliorated and what ranges are acceptable for the high scale threshold parameters.  All the while, it will be necessary to make sure that all of the physical scales in the problem like $M_{\rm Bulk}$ and $M_U$ are sensible.  For these reasons, we choose to work only with physical scales in what follows, leaving $M_{fict}$ behind.

\subsection{Exotics}
\label{sec:Exotics}

Now we finally come to the contributions from our new exotic fields.  As mentioned before, we can parametrize the exotic spectrum by the integers $K$, $L$, $M$, and $N$ as follows
\begin{equation}\begin{array}{c|c}
\text{Representation} & \text{Number} \\ \hline
(\mathbf{3},\mathbf{2})_{+1/6}\oplus (\mathbf{\overline{3}},\mathbf{2})_{-1/6} & M \\
(\mathbf{\overline{3}},\mathbf{1})_{-2/3}\oplus (\mathbf{3},\mathbf{1})_{+2/3} & M-N \\
(\mathbf{1},\mathbf{1})_{+1}\oplus (\mathbf{1},\mathbf{1})_{-1} & M+N \\ \hline
(\mathbf{\overline{3}},\mathbf{1})_{-1/3}\oplus (\mathbf{3},\mathbf{1})_{+1/3} & K \\
(\mathbf{1},\mathbf{2})_{-1/2}\oplus (\mathbf{1},\mathbf{2})_{+1/2} & K-L
\end{array}\label{exoticpar}\,,\end{equation}
where we must have
\begin{equation}M\ge |N|\qquad K>\text{min}(L,0)\end{equation}
in order for the parametrization to make sense.
Because we expect that the exotics will be very heavy, they will not run over a very large mass range.  Consequently, it should be sufficient to incorporate their effects on gauge coupling renormalization at only the 1-loop level.  The 1-loop $\beta$ function coefficients receive the following corrections from the exotic fields 
\begin{equation}\begin{split}
\delta b_1^{(\text{exotic})} &= 3M-\frac{2N}{5} + K - \frac{3L}{5} \\
\delta b_2^{(\text{exotic})} &= 3M+K-L \\
\delta b_3^{(\text{exotic})} &= 3M-N+K \,.
\end{split}\end{equation}
It is easy to verify that these corrections satisfy the Blumenhagen relation \eqref{blumenhagenrelation}.  Assuming all exotics get a mass at a common scale $M_{\rm Exotic}$ this leads to the corrections
\begin{equation}\delta_i^{(\text{exotics})} = \frac{\delta b_i^{(\text{exotic})}}{2\pi}\ln\left(\frac{M_{\rm Bulk}}{M_{\text{Exotic}}}\right)\,.\end{equation}
Note that, in keeping with the discussion of section \ref{subsec:highthresh}, the divergences from exotic running are capped off at the bulk ('winding') scale $M_{\rm Bulk}$ \cite{Conlon:2009xf,Conlon:2009qa}.  The contribution to $\Delta$ from the exotics and the shift in $M_U$ are simply
\begin{equation}\ba
\Delta^{(\text{exotics})} &= \frac{9}{14\pi}\left(L-N\right)\ln\left(\frac{M_{\rm Bulk}}{M_{\text{Exotic}}}\right)\cr
\delta\left(\frac{1}{2\pi}\ln\frac{M_U}{m_Z}\right)^{(\text{exotics})}  &= \frac{N-L}{28\pi}\ln\left(\frac{M_{\rm Bulk}}{M_{\text{Exotic}}}\right)\,.
\ea
\end{equation}
That these both vanish when $L=N$ reflects the fact that the $\beta$ function shifts are all universal in that case.  Only $\alpha_U$ would depend on such an overall shift.

\subsection{Summing Everything}

We are now ready to sum everything, including the high scale and exotic contributions we just discussed and the well-known corrections from  2-loop running and low scale thresholds that are reviewed in Appendix \ref{app:RG}.
The contributions to $\Delta$ and the shift in $M_U$ can be summarized as
\begin{equation}\begin{array}{c|cccc}
& \text{2-loop} & \text{Low Thresh} & \text{High Thresh} & \text{Exotic} \\ \hline
\Delta & \Delta^{(2-loop)}=-0.8197 & \frac{19}{28\pi}\ln\frac{m_{SUSY}}{m_Z} & \frac{3}{14\pi}\ln\frac{M_{\rm Bulk}}{M_U^{(0)}} +\Delta^{(finite)} & \frac{9}{14\pi}(L-N)\ln\frac{M_{\rm Bulk}}{M_{\text{Exotic}}} \\ \hline
\delta\left(\frac{1}{2\pi}\ln\frac{M_U}{m_Z}\right) & \delta_{M_U}^{(2-loop)}=0.07446 & -\frac{25}{168\pi}\ln\frac{m_{SUSY}}{m_Z} & \delta_{M_U}^{(finite)}
-\frac{19}{28\pi}\ln\frac{M_{\rm Bulk}}{M_U^{(0)}} & \frac{N-L}{28\pi}\ln\frac{M_{\rm Bulk}}{M_{\text{Exotic}}}
\end{array}\label{unisummary}\end{equation}
where we 
introduced notation for the 2-loop contributions $\Delta^{(2-loop)}$ and $\delta_{M_U}^{(2-loop)}$ so that we don't have to continually write the explicit numeric results in formulae.  The scale $m_{SUSY}$ is defined in \eqref{msusydef} and characterizes the low scale threshold corrections.


\subsubsection{Aside: Canceling Exotics and High Scale Thresholds?}
\label{subsubsec:exthresh}

Before turning to a general analysis, let us focus for a moment on the contributions from high scale thresholds and exotics.  The $\beta$ function shifts induced by exotics always satisfy the Blumenhagen relation \eqref{blumenhagenrelation} so the emergence of \eqref{blumenhagenrelation} from the IIB analysis of \cite{Blumenhagen:2008aw} hinted at an elegant scenario wherein the distortions of unification from exotics could precisely cancel those from KK threshold corrections \cite{Marsano:2009gv,Marsano:2009wr}.  From \eqref{unisummary} we see that exotics can cancel the logarithmic part of the KK corrections provided
\begin{equation}\left(\frac{M_{\rm Bulk}}{M_{\text{Exotic}}}\right)^{3(N-L)} = \frac{M_{\rm Bulk}}{M_U^{(0)}}\label{KKexoticcancel}\,.\end{equation}
This is consistent with the discussion of section \ref{subsubsec:blumenhagenrelation} since we expected there that a single pair of doublets, which corresponds to $N-L=-1$ in the parametrization \eqref{exoticpar}, could cancel the contribution from doublets in the $b_i^{(\text{Matter})}$ term in brackets in \eqref{introduceMfict} provided that they run over the energy range $(M_{fict}/M_{\rm Bulk})$.  The latter is simply $(M_{\rm Bulk}/M_U)^{1/3}$ from \eqref{Mfict}.

The condition \eqref{KKexoticcancel} is of course a more general one that applies to a generic exotic sector.  When $N-L>0$, an exact cancellation like this can never happen unless $M_{\text{Exotic}}>M_U^{(0)}$.  Since $M_U$ will not differ too much from $M_U^{(0)}$, it is very difficult to separate the scale $M_{\text{Exotic}}$ from the KK scale $M_U$ in this case.  When $N-L<0$, on the other hand, 
we are in even more trouble because \eqref{KKexoticcancel} can never be satisfied when $M_{\rm Bulk}$ is larger than both $M_{\text{Exotic}}$ and $M_U^{(0)}$.  Scenarios of the type proposed in \cite{Marsano:2009gv,Marsano:2009wr} that advocate cancellation of exotic splittings with those from logarithmic KK thresholds are therefore never realized.  The muddled mess of additional corrections in \eqref{unisummary} is therefore a welcome complication.

\subsubsection{Including all corrections}

The problems of section \ref{subsubsec:exthresh} really boil down to the fact that the exotic contribution to $\Delta$ tends to be larger in magnitude than the computable part of the high scale thresholds.  To deal with this, we need the remaining corrections to enter with the opposite sign of the exotic corrections.  Low scale thresholds tend to be rather small so the dominant effect that we can calculate is the one from 2-loop corrections $\Delta^{(2-loop)}$.  Negativity of $\Delta^{(2-loop)}$ means that it can help with the problems of section \ref{subsubsec:exthresh} if $N-L<0$ while it only makes matters worse when $N-L>0$.  This is somewhat unfortunate since we saw in section \ref{sec:surveysummary} that $N-L>0$ is preferred from the perspective of the $\mu$ problem.

It is perhaps more prudent to view the situation a little differently.  The MSSM has a built in problem with unification from 2-loop running.  In a generic scenario without exotics, one can try to tune high scale effects, encapsulated here by $\Delta^{(finite)}$ and $\delta_{M_U}^{(finite)}$, to compensate this.  Since exotics are forced on us, we can ask if their presence decreases or increases the required tuning.  When $N-L<0$ the exotics are helpful in general while $N-L>0$ leads to a situation in which they seem to be harmful.

How harmful are they?  Is the tuning of high scale parameters significantly worse than what we would need to compensate for 2-loop effects alone?  To address these questions, let us turn to a more detailed analysis.  We recall that our primary goals are as follows:
\begin{itemize}
\item Attain consistency with the observed low scale gauge couplings
\item Ensure that $M_{\rm Bulk}>M_U$
\item Achieve $M_{\text{Exotic}}< 10^{-2}M_U$ so that the exotics can be reliably separated from the KK tower
\end{itemize}
Our approach is to treat $M_{\rm Bulk}$, $M_{\text{Exotic}}$, and the finite threshold parameters $\Delta^{(finite)}$ and $\delta_{M_U}^{(finite)}$ as model parameters and adjust them as needed.  We start by making a choice for $M_{\text{Exotic}}$.  In a gauge mediation scenario of the type considered in the next section, this will (help) determine the sparticle masses from which we can directly compute $m_{SUSY}$.

We can then choose $M_{\rm Bulk}$ so that $\Delta=\Delta^{(ideal)}$.    This is done by taking
\begin{equation}\begin{split}\ln\frac{M_{\rm Bulk}}{M_U^{(0)}}&= \frac{1}{6(1-3(N-L))}\left[18(N-L)\ln\frac{M_U^{(0)}}{M_{\rm Exotic}}-19\ln\frac{m_{SUSY}}{m_z}\right. \\
&\qquad\qquad\qquad\qquad\qquad\left.+28\pi\left(\Delta^{(ideal)}-\Delta^{(2-loop)}-\Delta^{(finite)}\right)\right]\,.
\end{split}\end{equation}
With this choice, let us look at the structure of $\ln\frac{M_{\rm Bulk}}{M_U}$ and $\ln\frac{M_U}{M_{\text{Exotic}}}$.  We have
\begin{equation}\begin{split}\ln\frac{M_{\rm Bulk}}{M_U} &= \ln\frac{M_{\rm Bulk}}{M_U^{(0)}}+\ln\frac{M_U^{(0)}}{m_z}-\ln\frac{M_U}{m_z} \\
&= \ln\frac{M_{\rm Bulk}}{M_U^{(0)}} - 2\pi\,\,\delta\left(\frac{1}{2\pi}\ln\frac{M_U}{m_z}\right) \\
&= \frac{1}{18(N-L)-6}\left[-42(N-L)\ln\frac{M_U^{(0)}}{M_{\rm Exotic}}+(43+4[N-L])\ln\frac{m_{SUSY}}{m_Z}\right. \\
&\qquad\qquad\qquad\qquad + (66-2[N-L])\pi(\Delta^{(2-loop)}+\Delta^{(finite)}-\Delta^{(ideal)})\\
&\qquad\qquad\qquad\qquad\left.+(12-36[N-L])\pi(\delta_{M_U}^{(2-loop)}+\delta_{M_U}^{(finite)})\right] \\
\end{split}\end{equation}
\begin{equation}\begin{split}\ln\frac{M_U}{M_{\text{Exotic}}} &= \ln\frac{M_U}{m_z}-\ln\frac{M_U^{(0)}}{m_z}+\ln\frac{M_U^{(0)}}{M_{\text{Exotic}}} \\
&=2\pi\,\,\delta\left(\frac{1}{2\pi}\ln\frac{M_U}{m_z}\right) + \ln\frac{M_U^{(0)}}{M_{\text{Exotic}}} \\
&=\frac{1}{9(N-L)-3}\left[\left(21(N-L)-3\right)\ln\frac{M_U^{(0)}}{M_{\text{Exotic}}}-\left(12+2(N-L)\right)\ln\frac{m_{SUSY}}{m_z} \right. \\
&\qquad\qquad\qquad\qquad + ([N-L]-19)\pi\left(\Delta^{(2-loop)}+\Delta^{(finite)}-\Delta^{(ideal)}\right)\\
&\qquad\qquad\qquad\qquad \left. + (18[N-L]-6)\pi\left(\delta_{M_U}^{(2-loop)}+\delta_{M_U}^{(finite)}\right)\right]\\
\end{split}\end{equation}
We would like to use these relations to place bounds on the finite threshold parameters, for fixed $N$ and $L$, so that
\begin{equation}\ln\frac{M_{\rm Bulk}}{M_U}>0\qquad \ln\frac{M_U}{M_{\text{Exotic}}}>\ln 100\sim 4.6\label{desiredlns} \,.\end{equation}
These individual expressions are sufficiently complicated that it is difficult to learn anything by staring at them.  Perhaps more enlightening, however, is the result for $\ln M_{\rm Bulk}/M_{\text{Exotic}}$ which must also be larger than $\sim 4.6$ in order to have any hope of satisfying \eqref{desiredlns}
\begin{equation}\ln\frac{M_{\rm Bulk}}{M_{\text{Exotic}}} = \frac{1}{18(N-L)-6}\left[19\ln\frac{m_{SUSY}}{m_z}-6\ln\frac{M_U^{(0)}}{M_{\text{Exotic}}} + 28\pi(\Delta^{(2-loop)}+\Delta^{(finite)}-\Delta^{(ideal)})\right]\label{MBMExrat} \,.\end{equation}
This makes manifest the significant impact of the exotic fields and how quickly they become problematic as $|N-L|$  becomes large.  As $|N-L|$ grows, we very quickly need to compensate with hierarchically large values for the finite threshold correction $\Delta^{(finite)}$ of the sort that we find unacceptable.

\subsubsection{Application to Survey Models}

\begin{figure}
\begin{center}
\includegraphics[bb= 142 75 500 400,clip,width=0.5\textwidth]{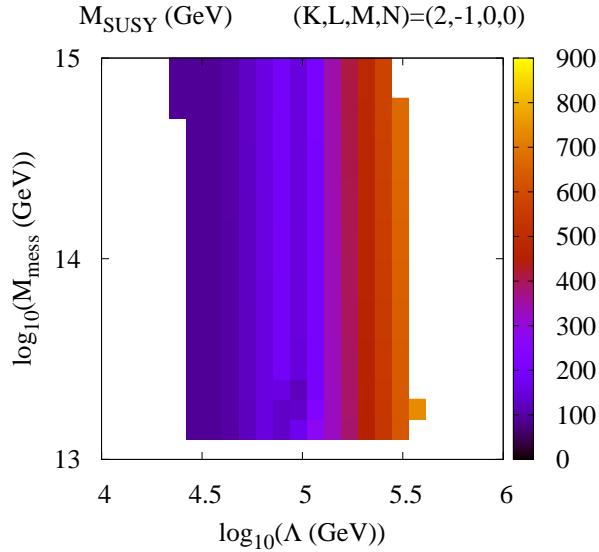}
\caption{This figure shows the effective mass scale $m_{SUSY}$ for low threshold corrections defined in the text in the $M_{\rm Mess}$ and $\Lambda$ plane. We have chosen the fluxes to be $(2,-1,0,0)$ so that $N-L=-1$, the same as for the benchmark point and scans discussed in Section~\ref{sec:phenomenology}.}
\label{fig:Msusy}
\end{center}
\end{figure}

Of the flux choices that arise in the survey of spectral cover construction in \cite{Dolan:2011iu}, the cases $N-L=\pm 2$ will be the most problematic for the unification story.
In figures \ref{fig:deltafinite2loop} and \ref{fig:deltafiniteMU}, we study the high scale threshold parameters $\Delta^{(finite)}$ and $\delta_{M_U}^{(finite)}$ for two benchmark choices of flux parameters $(K,L,M,N)=(0,-2,1,0)$ and $(3,2,0,0)$ that have $N-L=2$ and $-2$, respectively.

\begin{figure}
\begin{center}
  \twographs{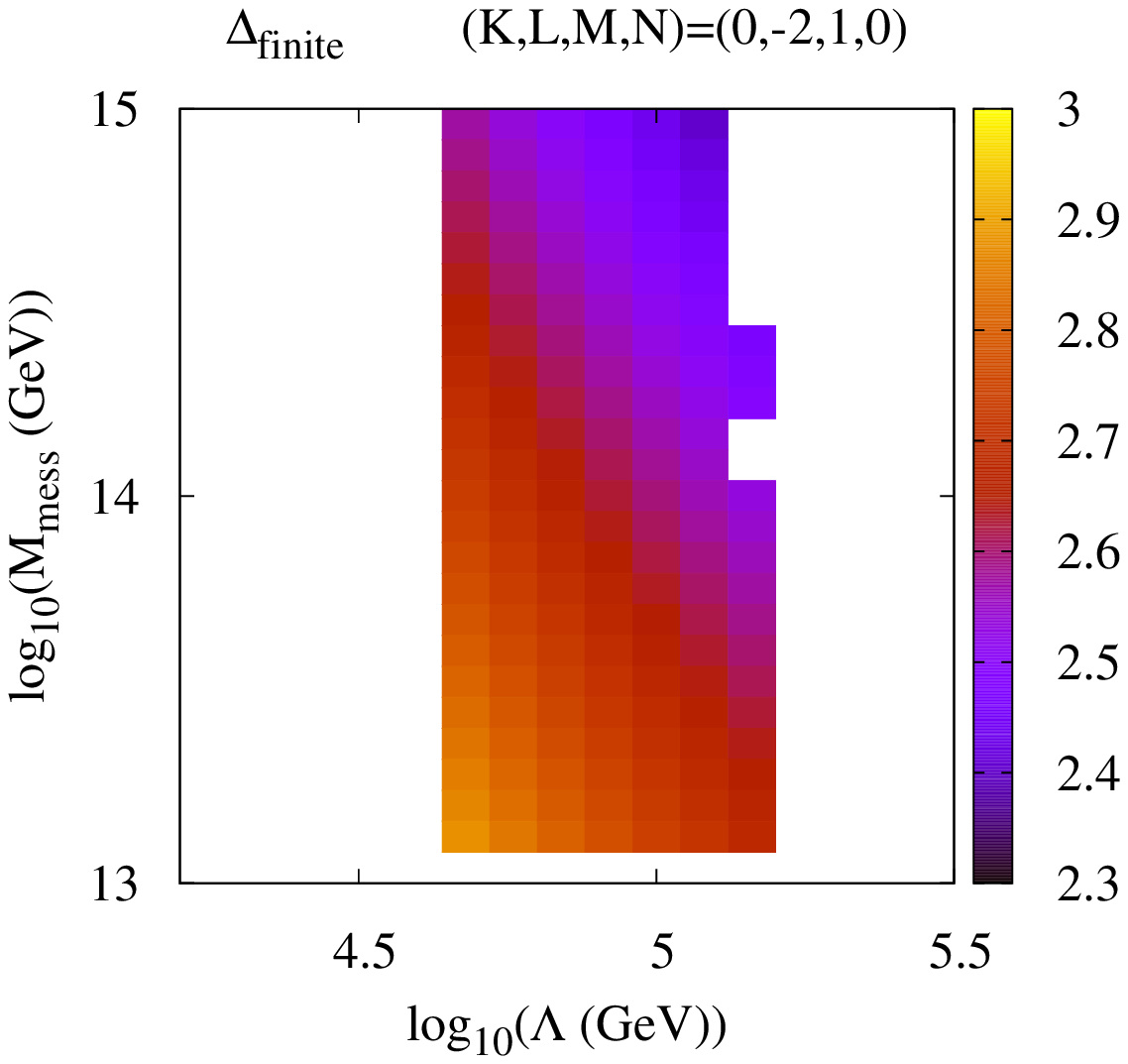}{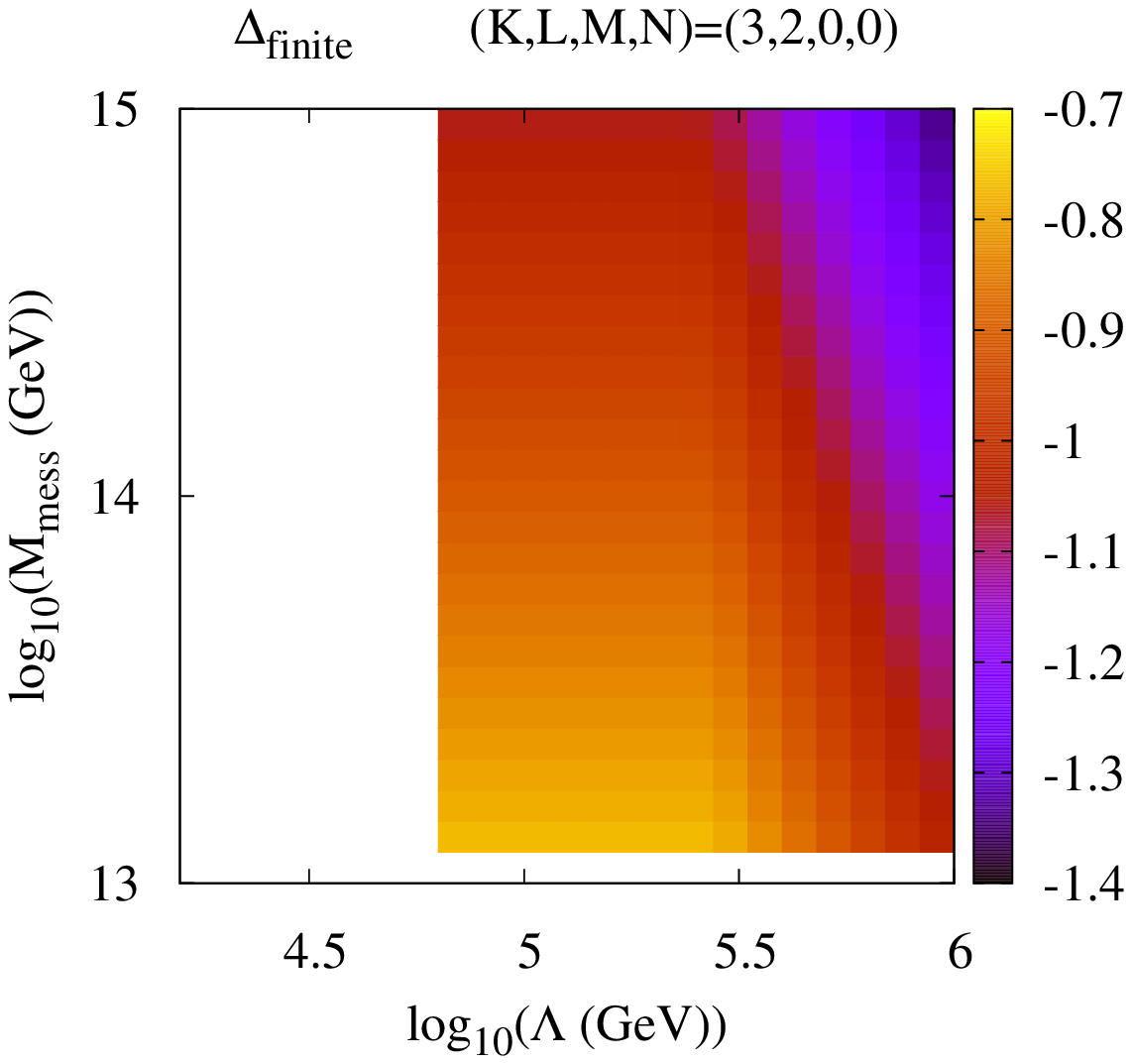}
\caption{Plots of $\Delta^{finite}$ in a parameter scan of (a) {{$(K,L,M,N)= (0,-2,1,0)$}} and (b) $(3,2,0,0)$ parameter spaces.}
\label{fig:deltafinite2loop}
\end{center}
\end{figure}

The first step to studying $\Delta^{(finite)}$ and $\delta_{M_U}^{(finite)}$ in each case is to compute $m_{SUSY}$.  For this, we assume a gauge mediation scenario of the type considered in the next section where the exotics play the dominant role in communicating supersymmetry breaking to the visible sector. 
As an example, we show in Figure~\ref{fig:Msusy} the results for $m_{SUSY}$ calculated in a scan over the resulting gauge mediated supersymmetry breaking (GMSB) parameter space with flux choice $(2,-1,0,0)$ which gives a maximally non-universal spectrum.  At the lowest values of the GMSB parameter $\Lambda$, $M_{SUSY}$ is equal to $M_Z$ and there are no low-scale threshold corrections. However, once LHC searches are taken into account, the minimum value of $m_{SUSY}$ possible is a few hundred GeV. We have computed the values of $m_{SUSY}$ for other flux choices as well, and find them generally to be a few hundred GeV in phenomenologically viable regions. An exception to this is the choice $(3,2,0,0)$, where $m_{SUSY}=m_Z$ over nearly the entire parameter space, rising to a maximum of 150~GeV.  In any case, $m_{SUSY}/m_Z$ is never very large so low scale threshold corrections tend to be rather small.

To study $\Delta^{(finite)}$ and $\delta_{M_U}^{(finite)}$ for our two benchmark flux choices $(0,-2,1,0)$ and $(3,2,0,0)$, we now scan over choices of $M_{\text{Exotic}}$ ranging from $10^{13}$ to $10^{15}$ GeV as well as the GMSB parameter $\Lambda$.  Within the scan, we impose the LEP bound on the Higgs mass so that points with $m_h<114.4$~GeV are not shown.  As we show in the next section, this is approximately the same effect that current LHC searches have.  For each choice of $M_{\text{Exotic}}$ and $M_{\rm Bulk}$, we choose $\delta_{M_U}^{(finite)}$ so that $M_{\rm Bulk}=M_U$ since this will allow us to maximize the separation between $M_U$ and $M_{\text{Exotic}}$.  We then fix $\Delta^{(finite)}$ so that $M_U=100M_{\text{Exotic}}$.  The results for $\Delta^{(finite)}$ are displayed in figure \ref{fig:deltafinite2loop} while those for $\delta_{M_U}^{(finite)}$ are shown in figure \ref{fig:deltafiniteMU}.  Note that the magnitude of $\Delta^{(finite)}$ is significantly larger when $N-L>0$ than when $N-L<0$ as our previous discussion suggests.  When $N-L>0$, the exotic contribution adds to the 2-loop one and $\Delta^{(finite)}$ must be quite large to compensate both.  When $N-L<0$, the exotic contribution competes with the 2-loop one and $\Delta^{(finite)}$ can be smaller. For part of the $N-L=-2$ parameter range, we even have that $|\Delta^{(finite)}|<|\Delta^{(2-loop)}|$ so that the tuning of high scale effects that we must assume to attain $\Delta=\Delta^{(ideal)}$ can be less than in a generic extension of the MSSM.  More generally, we see from both cases that $|N-L|=2$ is small enough that the high scale threshold parameters do not differ much from ${\cal{O}}(1)$.  The tuning of high scale effects is not improved relative to the usual 2-loop problems of the MSSM but it isn't significantly worse either.

\begin{figure}
\begin{center}
  \twographs{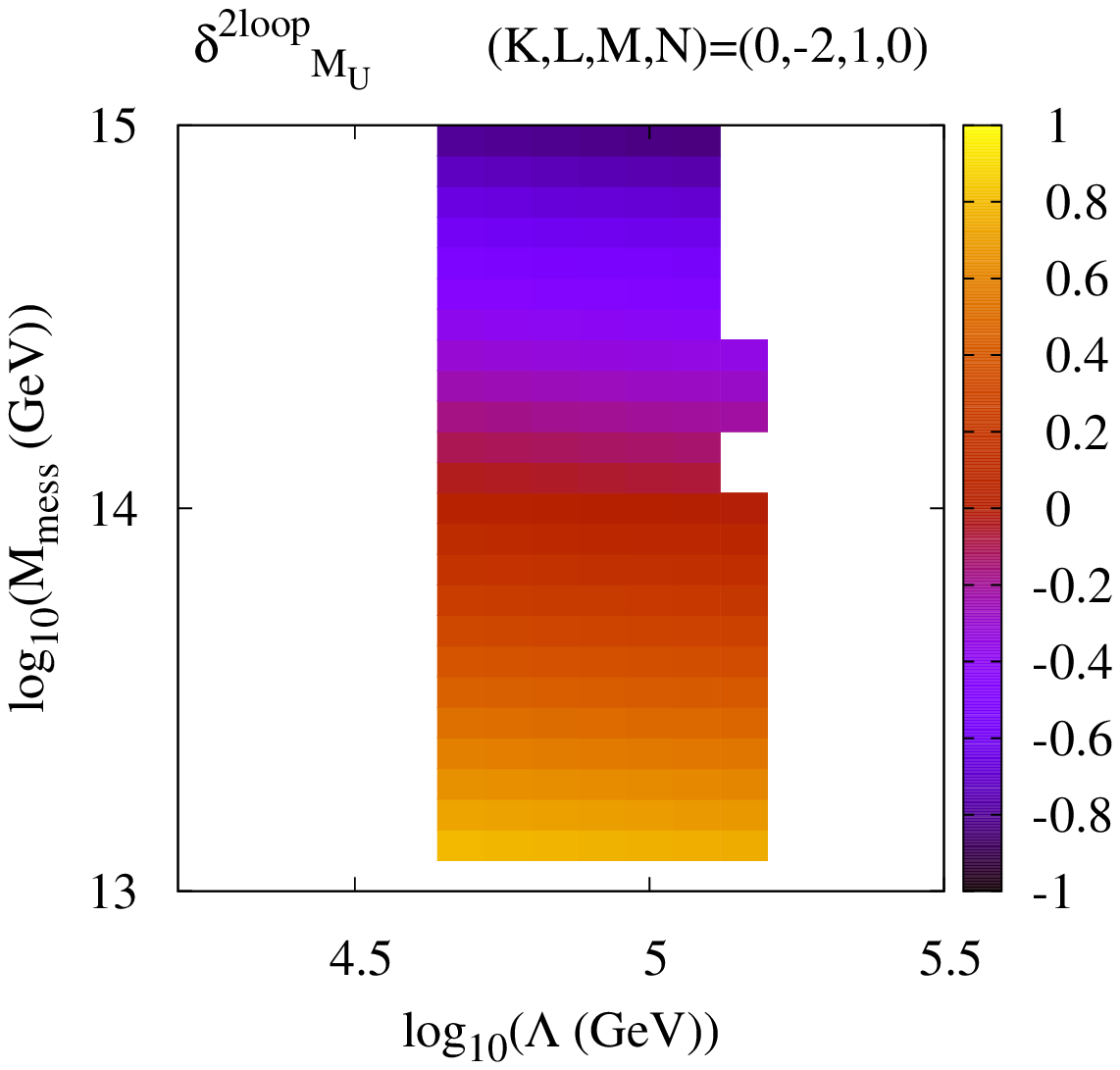}{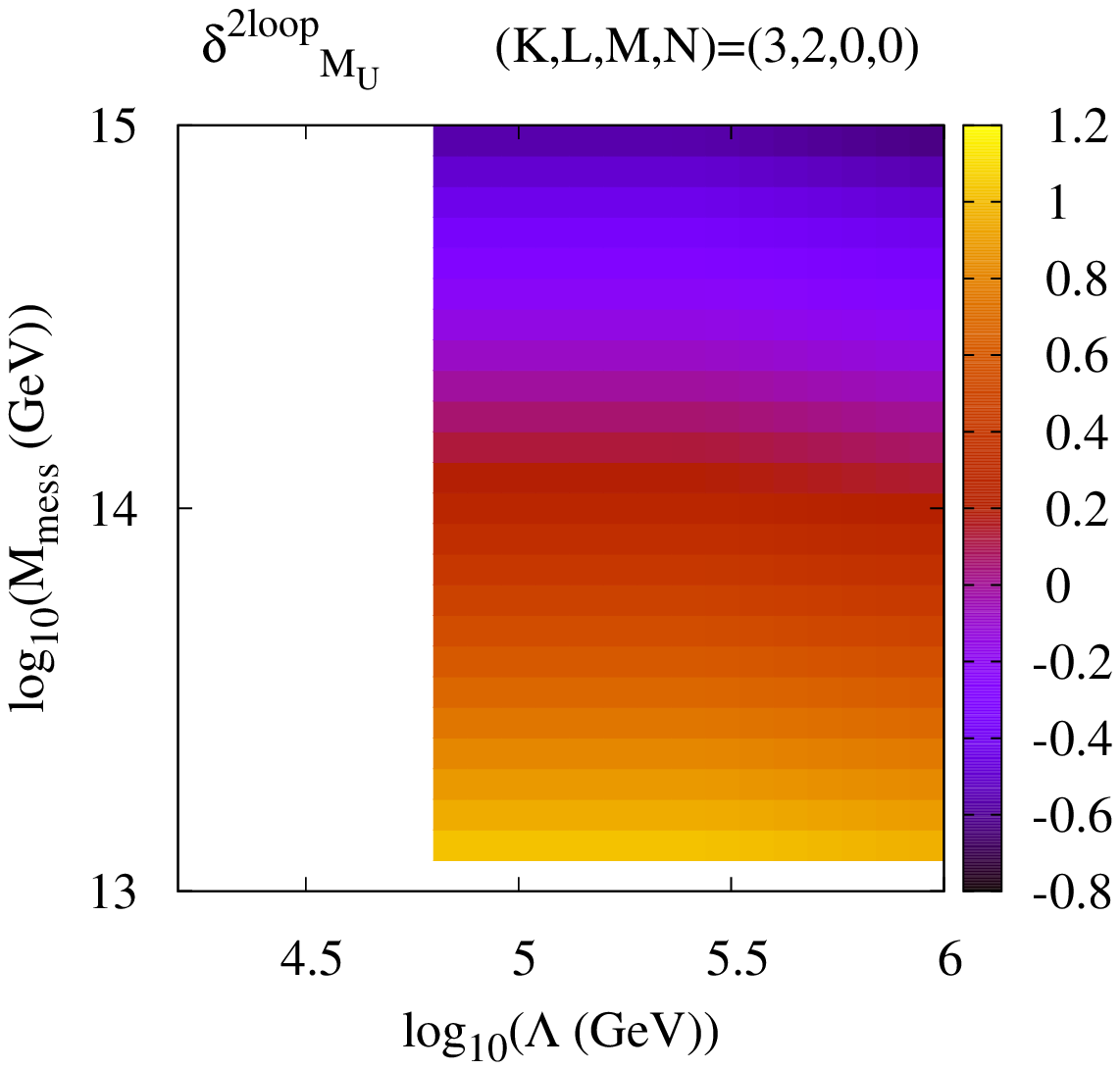}
\caption{Plots of $\delta^{finite}_{M_U}$ in a parameter scan of (a)  {{$(K,L,M,N)= (0,-2,1,0)$}} and (b) $(3,2,0,0)$ parameter spaces.}
\label{fig:deltafiniteMU}
\end{center}
\end{figure}

\newpage

\section{Gauge Mediation and Phenomenology}
\label{sec:phenomenology}

\subsection{Gauge Mediation with non-GUT messengers}

{{The coupling \eqref{Xffb} of exotic fields to the MSSM singlet $X$ 
\begin{equation}W\sim X f_{ex}\overline{f}_{ex}\label{GMSBcoup}\end{equation}
is not only necessary to lift them, it is also highly suggestive.  As it does not carry MSSM charge, $X$ is associated to a curve on the internal space that extends away from the GUT 7-branes into the bulk of the compactification.  If we suppose that supersymmetry breaking originates somewhere in the bulk, perhaps from a hidden sector on some distant stack of branes, it is quite plausible that $X$ will be sensitive to it and acquire an $F$-component expectation value.  Through the coupling \eqref{GMSBcoup}, then, our exotics become a vehicle for communicating supersymmetry breaking to the visible sector.  If we further assume that exotics provide the dominant method of communication, we arrive at a very odd but distinctive scenario for gauge mediated supersymmetry breaking (GMSB) in which the messenger fields do not comprise complete GUT multiplets{\footnote{The assumption that gauge mediated contributions dominate requires a relatively small gravitino mass.  In string models where moduli stabilization can be achieved in a supergravity setting, the recent work of \cite{Acharya:2010af} indicates that this may not be particularly natural.}}.  The rest of this paper is devoted to a study of this type of GMSB scenario.

The natural appearance of structures that suggest GMSB was noted in early studies of F-theory GUTs \cite{Marsano:2008jq,Heckman:2008qt}.  There, the focus was on scenarios of the type advocated in \cite{Ibe:2007km} where the $\mu$-term is dynamically generated by a Giudice-Masiero type mechanism \cite{Giudice:1988yz} from
\begin{equation}\frac{1}{M_U}\int\,d^4\theta\,X^{\dag}H\overline{H}\end{equation}
or some suitable operator of higher dimension that reduces to a $\mu$ term in the presence of bosonic and $F$-component expectation values for $X$.  In our models, which make the simplifying assumption that only one singlet $X$ picks up expectation values of any type, $\mu$ is generated in this way when $N-L>0$.  When $N-L<0$, it is generated instead through the bosonic expectation value of $X$ via an operator of the form \eqref{regrownoperators}.  Since we will consider benchmark studies of exotic/messenger sectors corresponding to all values of $N-L$ for completeness, it is important to keep in mind that the $N-L<0$ models will require significant tuning or additional structure to address the $\mu$ problem.   
}}

{{ We turn now to the basic structure of our gauge mediated scenarios.}} The leading gauge mediated soft masses that one obtains at the messenger scale $M_{\rm Mess}{= M_{\rm Exotic}}$ take the standard form, in particular the one-loop gaugino masses are 
\begin{equation}M_{1/2}(M_{\text{Mess}}) =  \delta b_i \left(\frac{\alpha_i(M_{\text{Mess}})}{4\pi}\right)\left(\frac{F}{M_{\text{Mess}}}\right)\,,
\end{equation}
and  the 2-loop squark and slepton masses are given by
\begin{equation}m^2_Q(M_{\text{Mess}})=\sum_{\text{relevant }i}\frac{c_i\,\delta b_i\,\alpha_i(\mu)^2}{8\pi^2}\left|\frac{F}{M_{\text{Mess}}}\right|^2\,.
\end{equation}
In these formulae, the $\delta b_i$ denote shifts of the $\beta$ function coefficients  induced by the messenger sector, the $c_i$ denote quadratic Casimirs of the MSSM gauge groups, and the dimensionful quantities $F$ and $M_{\text{Mess}}$ arise as the bosonic and $F$-component expectation values of the singlet field $X$. 
{{The $\beta$ function shifts from our exotic fields were determined in section \ref{sec:Exotics} to be}}
\begin{equation}\begin{split}
\delta b_1 &= 3M-\frac{2}{5}N + K - \frac{3}{5}L\\
\delta b_2 &= 3M + K -L \\
\delta b_3 &= 3M - N  + K \,.
\end{split}\end{equation}
The flux parameters are subject to the relations (\ref{exoticparam}). 
The trilinear $A$-terms and $B_{\mu}$ term are all zero at the messenger scale.
Notice that the beta-function shifts are all equivalent provided $L=N$.  {{This is not achievable in F-theory GUT models with $U(1)_{PQ}$ on general grounds \cite{Marsano:2010sq,Dolan:2011iu}.}}

\subsection{Benchmark Points}

In this section we discuss some of the phenomenology of our model.  To generate these points we employ a modified version of \texttt{SoftSusy 3.0.13}~\cite{Allanach:2001kg} which takes $B$ instead of $\tan\beta$ as an input and uses the electroweak symmetry breaking conditions to set $\tan\beta$ and $\mu$ at the weak scale. As discussed in~\cite{Rattazzi:1996fb,Babu:1996jf,Abel:2009ve,Abel:2010vba} this generally leads to large values of $\tan\beta$ {in mGMSB and other universal models}. 
Points with a tachyon in the spectrum, with a Higgs potential unbounded from below, which do not break electroweak symmetry or which have couplings that become non-perturbative before the messenger scale are rejected. The lightest Higgs boson is SM-like, and so we apply the LEP bound of $m_h > 114.4$~GeV. This bound imposes a strong restriction on the SUSY spectrum. In general, it is possible to have relatively light supersymmetric particles and a Higgs boson whose mass exceeds the LEP bound if the trilinear soft supersymmetry breaking couplings are sizable and negative. In gauge mediated scenarios this is not the case, as the trilinear coupling are generically induced at the two-loop level, and are hence small. Therefore gauge mediation leads to a heavier superparticle spectrum than other mediation scenarios which can accommodate larger values of the trilinear soft terms. Once the LEP bound is taken into account, the majority of the remaining parameter space evades the current (\textit{circa} $1fb^{-1}$) searches from the LHC.

The shifts in the beta-functions lead to gaugino non-universality at the GUT scale. How large can this non-universality get, and to what choice of fluxes does it correspond? In ordinary gauge mediation and mSUGRA the gaugino mass parameters at the low scale are given in the ratios
\begin{equation}
M_1 : M_2 : M_3 \simeq 1:2:6 = \alpha_1 : \alpha_2 : \alpha_3 \,.
\label{eq:GUTgauginos}
\end{equation}
With general values of the flux parameters the gaugino masses at the low scale in our model are given by 
\be
M_1 : M_2: M_3 \simeq  1: 2 \left(\frac{3M+K-L}{3M- 2N/5 +K - 3L/5}\right) : 6 \left( \frac{3M+K-N}{3M+K-2N/5 -3L/5} \right) \,.
\label{eq:nonGUTgauginos}
\ee
For each of the three values of $\Delta= -2, 1, 2$ in (\ref{DeltaVals}) that are part of the survey, we will now figure out how non-universality can be maximized. 
{{Furthermore, for simplicity, we shall set
\be
N=0 \,.
\ee
}}

\subsubsection{$L =-1$}

We require
\begin{equation}M,K\ge 0\,.\end{equation}
The gaugino mass ratios become
\begin{equation}\frac{M_2}{M_1}=2\left(1+\frac{2}{5(3M+K+3/5)}\right)\qquad\frac{M_3}{M_1}=6\left(1-\frac{3}{5(3M+K+3/5)}\right) \,.\end{equation}
Nonuniversality is maximized when $3M+K$ is as small as possible.  This makes sense intuitively.  By fixing $L$ we are fixing the nonuniversal contributions since we will only have one excess pair of doublets in addition to complete GUT multiplets in the messenger sector.  If we crank up $M$ and $K$ we increase the universal contributions so that they dominate the nonuniversal ones.  The most interesting models then are the ones with small $M$ and $K$ which is to say the `minimal' ones without a lot of extra messengers.

The most extreme possibility is $M=0$ and $K=1$.  Here we find
\begin{equation}\frac{M_2}{M_1} = 2\times \frac{5}{4}\qquad \frac{M_3}{M_1}=6\times \frac{15}{24}\qquad \frac{M_3}{M_2}= 3\times \frac{1}{2}.\end{equation} We have found that the $M=0$, $K=1$ parameter space is already severely constrained by LHC searches for jets plus missing energy at low masses. At higher masses the parameter space quickly runs out due to the large values of $\tan\beta$ predicted in this scenario (greater than fifty), leading to problems with non-pertubativity and unstable vacua. We therefore choose to focus on the next-most non-universal scenario, $M=0$, $K=2$ for our phenomenological studies.
To see how quickly things become universal consider $M=0$, $K=2$ and $M=K=1$.  We have
\begin{equation}\begin{array}{ccc|ccc}
L & M & K & (1/2)\times M_2/M_1 & (1/6)\times M_3/M_1 & (1/3)\times M_3/M_2 \\ \hline
0 & M & K & 1 & 1 & 1 \\ \hline
-1 & 0 & 1 & 5/4\sim 1.25 & 5/8\sim 0.63 & 1/2\sim 0.5 \\
-1 & 0 & 2 & 15/13\sim 1.15 & 10/13\sim 0.77 & 2/3\sim 0.67 \\
-1 & 1 & 0 & 10/9\sim 1.11 & 5/6\sim 0.83 & 3/4=0.75 \\
-1 & 1 & 1 & 25/23\sim 1.09 & 20/23\sim 0.87 & 4/5\sim 0.80
\end{array}\end{equation}

We can now try to look for a large $M,K$ benchmark.  We check perturbativity by requiring that $\alpha_i^{-1}$ are all $\ge 0$ at $2\times 10^{16}$ GeV.  For each $i$, we solve for the value of $K$ as a function of $M$ that sets $\alpha_i^{-1}=0$ at $2\times 10^{16}$ GeV.  We then plug back into the other $\alpha_j$'s to see if they are all still positive.  In this way we determine that $\alpha_2^{-1}$ is the first coupling to pass through zero at $2\times 10^{16}$ GeV as we increase $K$ and $M$.

Let us now set $\alpha_2^{-1}$ to zero at $2\times 10^{16}$ GeV.  This gives us a relation
\begin{equation}3M+K<28 \,.\end{equation}
To maximize this let us set
\begin{equation}K=27-3M \,.\end{equation}
The gaugino mass ratios do not depend on the particular choice of $M$ in this case.  We find
\begin{equation}\begin{array}{ccc|ccc}
L & M & K & (1/2)\times M_2/M_1 & (1/6)\times M_3/M_1 & (1/3)\times M_3/M_2 \\ \hline
0 & M & K & 1 & 1 & 1 \\ \hline
-1 & M & 27-3M & 70/69\sim 1.01 & 45/46\sim 0.98 & 27/28\sim 0.96
\end{array}\end{equation}

\subsubsection{$L =-2$}

We require
\begin{equation}M\ge K\ge 0\,.\end{equation}
The gaugino mass ratios become
\begin{equation}\frac{M_2}{M_1}=2\left(1+\frac{4}{5(3M+K+6/5)}\right)\qquad\frac{M_3}{M_1}=6\left(1-\frac{6}{5(3M+K+6/5)}\right)\,.\end{equation}
Again, deviation from universality is achieved by minimizing $3M+K$.  Here, the minimal values are $M=1$, $K=0$.  We can tabulate a few things in the neighborhood of that
\begin{equation}\begin{array}{ccc|ccc}
L & M & K & (1/2)\times M_2/M_1 & (1/6)\times M_3/M_1 & (1/3)\times M_3/M_2 \\ \hline
0 & M & K & 1 & 1 & 1 \\ \hline
-2 & 1 & 0 & 25/21\sim 1.19 & 5/7\sim 0.71 & 3/5=0.6 \\
-2 & 1 & 1 & 15/13\sim 1.15 & 10/13\sim 0.77 & 2/3\sim 0.67 \\
-2 & 1 & 2 & 35/31\sim 1.13 & 25/31\sim 0.81 & 5/7\sim 0.71 \\
-2 & 2 & 0 & 10/9\sim 1.11 & 5/6\sim 0.83 & 3/4\sim 0.75 \\
\end{array}\end{equation}

\subsubsection{$L =2$}

In this case we are forced to take
\begin{equation}M=0\,,\qquad K\ge 2\,.\end{equation}
The gaugino mass ratios become
\begin{equation}\frac{M_2}{M_1}=2\left(1-\frac{4}{5K-6}\right)\qquad \frac{M_3}{M_1}=6\left(1+\frac{6}{5K-6}\right)\,.\end{equation}
Clearly $K=2$ will give the maximal deviation from nonuniversality.  $K=2$ is a somewhat degenerate case, though, because the 1-loop contribution to $M_2$ vanishes.  This is easy to understand because $K=L=2$ corresponds to having a vector-like pair of triplet messengers and no doublet messengers.  In this case, $M_2$ will get its dominant contribution at 2-loops.  This could potentially be very interesting.  For now, though, let us throw out this case and look at larger values of $K$.  We tabulate results for a few small values of $K$.

\begin{equation}\begin{array}{ccc|ccc}
L & M & K & (1/2)\times M_2/M_1 & (1/6)\times M_3/M_1 & (1/3)\times M_3/M_2 \\ \hline
0 & M & K & 1 & 1 & 1 \\ \hline
2 & 0 & 3 & 5/9\sim 0.56 & 5/3\sim 1.67 & 3 \\
2 & 0 & 4 & 5/7\sim 0.71 & 10/7\sim 1.43 & 2 \\
2 & 0 & 5 & 15/19\sim 0.79 & 25/19\sim 1.32 & 5/3\sim 1.67
\end{array}\end{equation}

\subsubsection{Benchmark Models}\label{subsubsec:Bench}

In summary, we shall consider benchmark choices for the flux numbers $K,L,M,N$, where $N=0$ and which probe the regimes far and  close to universality of the gaugino masses, as well as, for comparison, mGMSB with ${\bf 5}$ messengers. The choices of benchmark points are shown in the table below, along with a `lookalike' point from mGMSB which has been selected to mimic (as much as possible) the phenomenology of the $\hbox{NonUniv}_{L=-1}^{\rm{SmallFlux}}$ point.  In the following sections we explore the parameter spaces and phenomenology of these flux choices. 

\be\label{BenchMarkPoints}
\begin{array}{|c|l|c|c|c|c|c|}\hline
  L &\hbox{Benchmark Model}  & (K,\, L,\, M,\, N)  & M_{\rm Mess}  & \Lambda & m_{\tilde{g}} \cr \hline
 -2&  \hbox{NonUniv}_{L=-2}&  (0,-2,1,0)  &  10^{14} \text{GeV} & 8 \times 10^{4}\text{GeV} & 1686\text{GeV}\cr
 -1& \hbox{NonUniv}_{L=-1}^{\rm{SmallFlux}} &    (2,-1,0,0)   &    10^{14} \text{GeV}  & 9 \times 10^4\text{GeV} & 1307\text{GeV}  \cr
  +2& \hbox{NonUniv}_{L=2} &    (3,2,0,0) &  10^{14}\text{GeV} & 7\times 10^{4}\text{GeV} & 1489\text{GeV}\cr
 -1 & \hbox{NonUniv}_{L=-1}^{\rm{LargeFlux}}&   (6,-1,7,0)  &  10^{14}  \text{GeV}& 1.3\times 10^4 \text{GeV} &  2337\text{GeV}\cr
\hline\hline
0&\hbox{mGMSB5}	   &  (3,0,0,0)  &  10^{10}\text{GeV} & 7\times 10^{4}\text{GeV} &  1503\text{GeV} \cr\hline
\end{array}
\ee

\subsection{Parameter Space and Spectra}

Unification favors for each of the non-universal benchmark points in section \ref{subsubsec:Bench}
a range for the messenger scale $M_{\rm Mess}$. Within this range, we will now scan the
$M_{\rm Mess}-\Lambda$ plane, where 
\be
\Lambda = {F\over M_{\rm Mess}} \,,
\ee
provide benchmark spectra and mGMSB lookalike models.
The range of gluino masses that can be achieved in the benchmark flux choices, while being compatible with the LEP bound and for fixed $M_{\rm Mess} = 10^{14}$~GeV, is bounded below by the following values

\be
\begin{array}{|c||c|c|c|c|}\hline
 (K,\, L,\, M,\, N)  & (0,-2,1,0) &  (2,-1,0,0) & (3,2,0,0) &  (6,-1,7,0)   \cr\hline
  m_{\tilde{g}}^{\rm{min}} &     940\text{GeV} & 945\text{GeV} & 1370\text{GeV} &  1000\text{GeV}
  \cr \hline
\end{array}
\ee
{The $m_{\tilde{g}}^{\rm{min}}$ value for the $(3,2,0,0)$ benchmark is significantly larger than the other models due to the low values of $\tan\beta$ (less than 10) in this model. }
For the actual benchmark points (\ref{BenchMarkPoints}) we have chosen slightly higher values of the gluino mass, to evade any immediate ruling out of the models by the LHC. {However these lower bounds are still very similar to the recent limits from the ATLAS and CMS collaborations. Therefore, once the LEP bound is taken into account it is not surprising that there is no sign of SUSY yet.}


\subsubsection{$\hbox{NonUniv}_{L=-1}^{\rm{SmallFlux}}$  and mGMSB Lookalike}

The small flux model $\hbox{NonUniv}_{L=-1}^{\rm{SmallFlux}}$  has $(K,L,M,N)=(2,-1,0,0)$.
Figure~\ref{fig:tanbplot1} (a) shows $\tan\beta$ in the $M_{\rm Mess}-\Lambda$ plane, with logarithmic axes. This plot 
does not include the cuts from the sparticle direct search constraints, and whether a point from the scan is included is only determined by whether there are no tachyons or other problems\footnote{{All spectrum plots in this paper have been made using the PySLHA package.}}. Three contours are also shown. The blue dashed one is a gluino mass contour of 1~TeV, which represents the approximate effect off the LHC searches for supersymmetric particles. The black contour is the LEP bound on the mass of the Higgs, $m_h = 114.4$~GeV, and the green dashed contour delineates the identity of the NLSP. To the right of the green contour the NLSP is the lightest stau and to the left it is the bino-like lightest neutralino. This set of fluxes leads to large values of $\tan\beta$, between 40 and 50 once the LEP limit is taken into account.

\begin{figure}
\begin{center}
\twographs{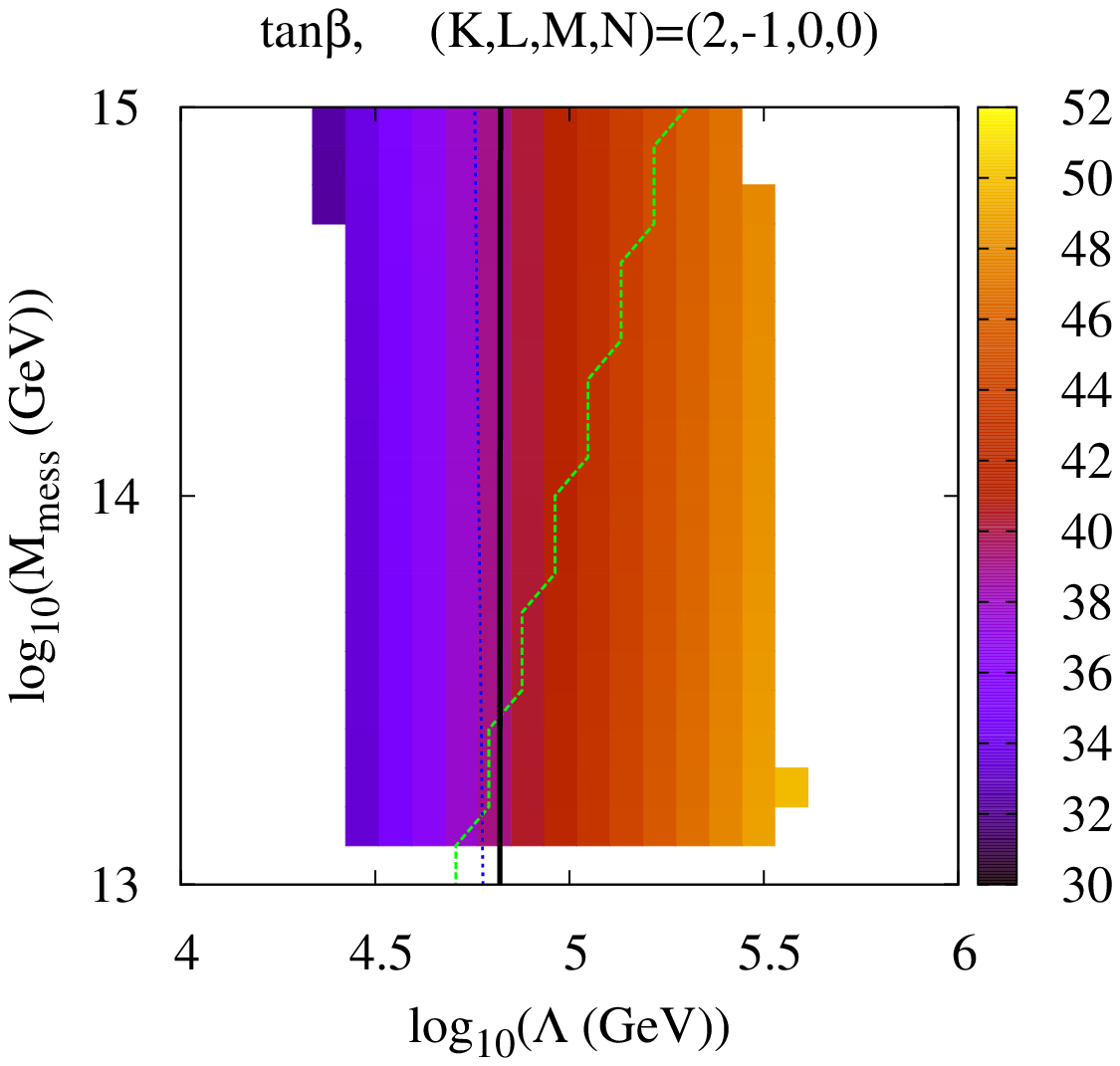}{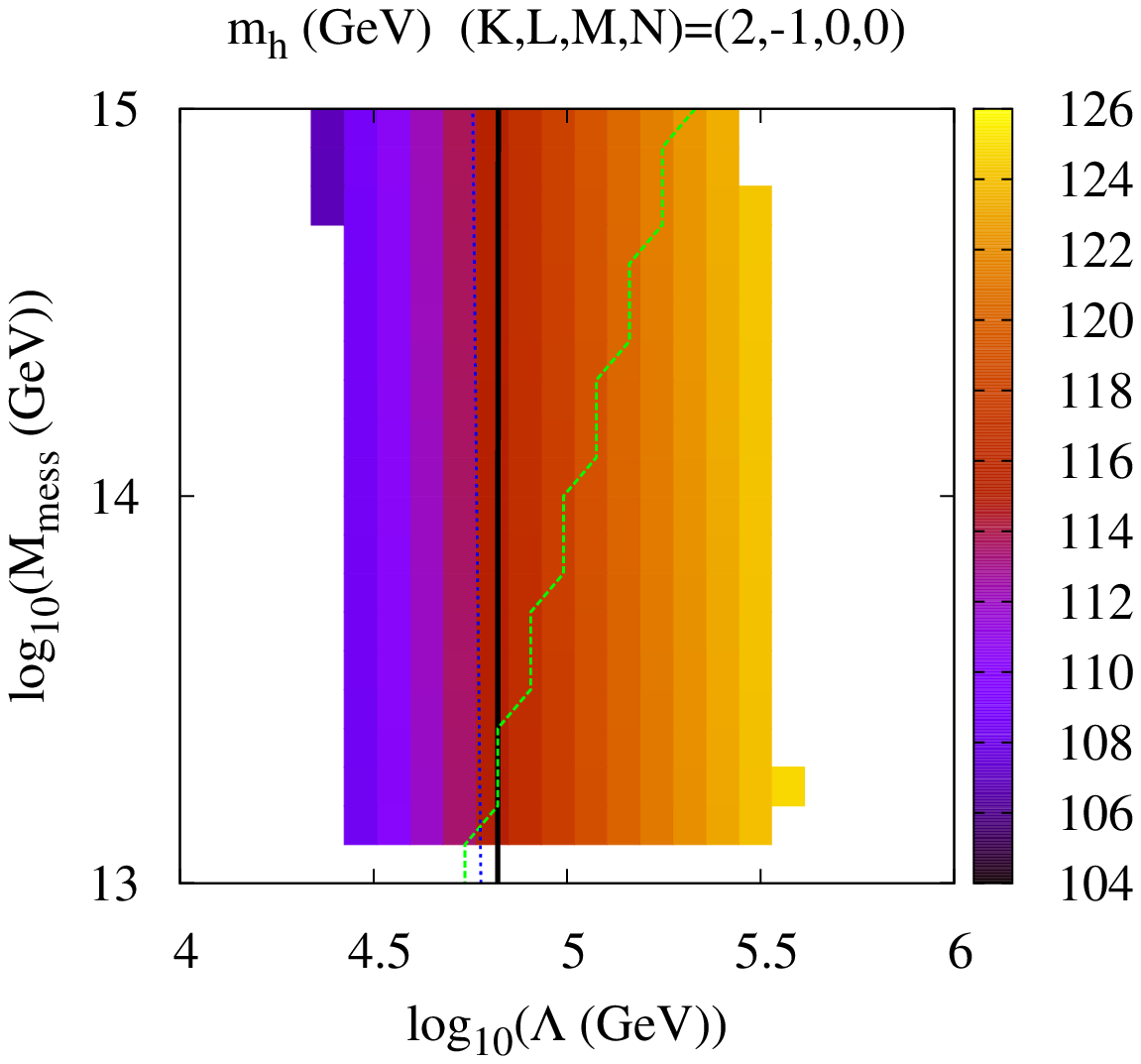}
\caption{The left-hand plot (a) shows the viable parameter range in the $\log_{10}(M_{\rm Mess})-\log_{10}(\Lambda)$ plane for the flux choice $(2, -1, 0, 0)$, i.e. the model $\hbox{NonUniv}_{L=-1}^{\rm{SmallFlux}}$. It is coloured according to the values of $\tan\beta$ obtained at each viable point. The right hand plot (b) shows the mass of the lightest Higgs boson $h^0$. The contour line in this case is the exclusion limit from LEP, $m_h > 114.4$~GeV.}
\label{fig:tanbplot1}
\end{center}
\end{figure}

In Figure~\ref{fig:tanbplot1} (b) we show the mass of the lightest CP even Higgs boson $m_h$ in the $M_{\rm Mess}-\Lambda$ plane with logarithmic axes. The contour lines are the same as shown in Figure~\ref{fig:tanbplot1} (a). Of note is that the Higgs bound from LEP (the black line) lies above the 1~TeV gluino mass contour (the dashed blue line). In that sense, it is therefore no surprise that no signals of supersymmetry have been observed so far. 
 This particular choice of fluxes leads to the coloured and weakly interacting parts of the spectrum becoming more compressed than usual, by a factor of 20-30\%. However, the hierarchy between the weak $SU(2)$ gauginos and the bino is increased by 15\%.

\begin{figure}
\begin{center}
\epsfig{file=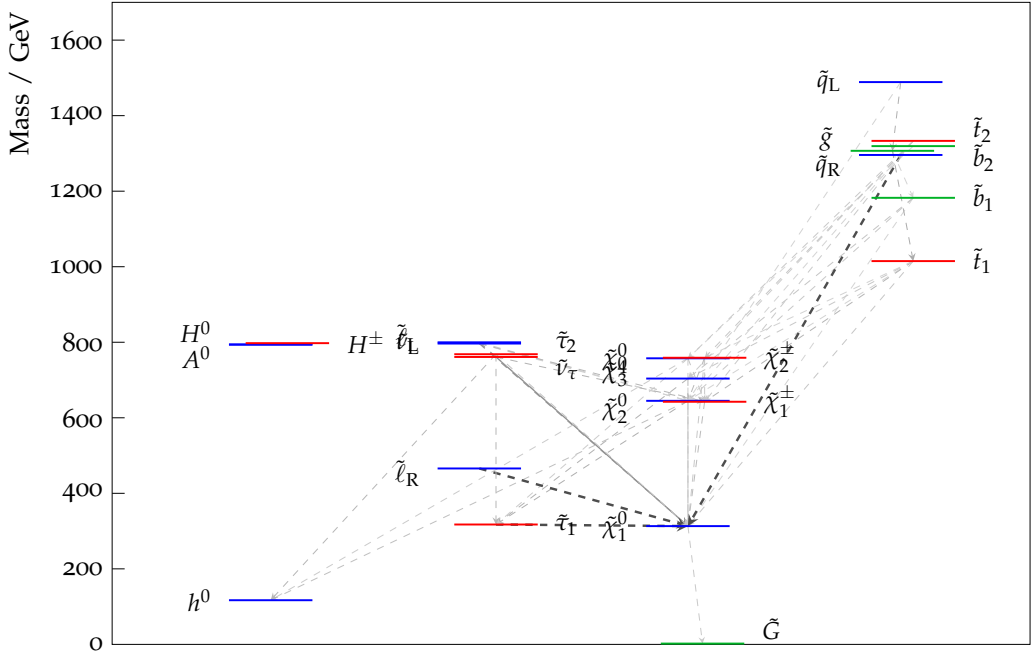,width=0.8\textwidth}
\caption{This figure shows the spectrum of the benchmark model  $\hbox{NonUniv}_{L=-1}^{\rm{SmallFlux}}$ with $M_{\rm Mess} = 10^{14}$~GeV and $\Lambda = 9 \times 10^4$~GeV.}
\label{fig:M1N1Spec}
\end{center}
\end{figure}

\begin{figure}
\begin{center}
\epsfig{file=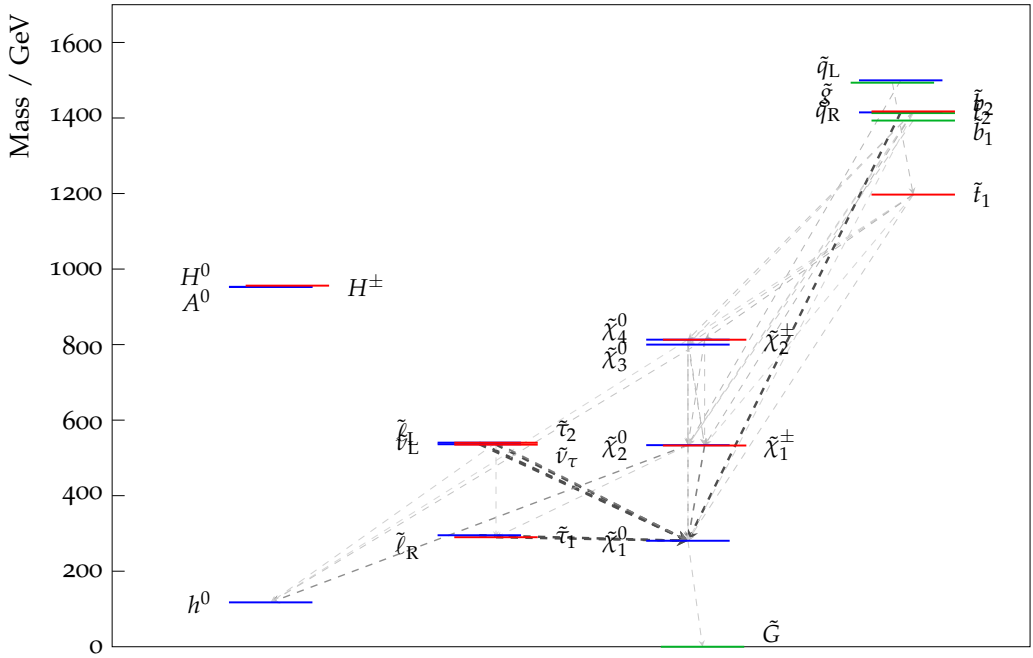,width=0.8\textwidth}
\caption{This figure shows the spectrum of an mGMSB point with similar phenomenology to  $\hbox{NonUniv}_{L=-1}^{\rm{SmallFlux}}$.}
\label{fig:mGMSBspec1}
\end{center}
\end{figure}

In Figure~\ref{fig:M1N1Spec} we show a benchmark spectrum of one of the points in the small flux parameter space. We have chosen the theoretically motivated value $M_{\rm Mess}= 1\times 10^{14}$~GeV, the value of $\Lambda$ is $9\times 10^4$~GeV and we obtain $\tan\beta= 41.7$. 

This point has neutralino NLSP, which is predominantly bino-like. The right handed sleptons are slightly heavier than the neutralino, with the lightest stau nearly being a co-NLSP to the neutralino, although for this point the stau-neutralino splitting for this point is greater than the $\tau$ mass and so the stau should decay promptly. The large values of $\tan\beta$ characteristic of gauge mediated models with $B_{\mu}=0$ lead to large splittings between the slepton masses.

The left handed sleptons are slightly heavier than the remaining weakly interacting gauginos, and the masses of the strongly interacting sparticles are all above 1~TeV. The lightest strongly interacting particle is the lightest stop, which is as usual somewhat lighter than the other squarks due to the mixing effects and the large value of the top Yukawa coupling. An important question is how distinguishable is our model from a benchmark model such as mGMSB. We will return to this later on, but note here that it is not easy for a mGMSB point to completely mimic the benchmark point we have chosen. 
The shifts in the beta-functions induced by the fluxes is $(\delta b_1, \delta b_2, \delta b_3 )= (2.6,3,2)$. This choice of fluxes therefore interpolates between mGMSB with 2 and 3 three sets of $5 \oplus \bar{5}$ messengers. Thus while in the weak sector we can find good agreement for $N_5=3$, the strong sector is a better fit for $N_5=2$.


\subsubsection{$\hbox{NonUniv}_{L=-1}^{\rm{LargeFlux}}$  and mGMSB Lookalike}

We now turn to the large flux models $ \hbox{NonUniv}_{L=-1}^{\rm{LargeFlux}}$ with $(K,L,M,N)=(6,-1,7,0)$.
In this regime of parameter space the NLSP is the stau. This is for two reasons. One is because in mGMSB $m_{1/2}/m_0$ scales as $\sqrt{N_{mess}}$ so that increasing the number of effective messengers (i.e. turning up the fluxes) increases the masses of the gauginos relative to the scalars. The second is that the stau mixing is proportional to $\tan\beta$, and the right-handed stau can become quite light when $\tan\beta$ is large.

\begin{figure}
\begin{center}
\twographs{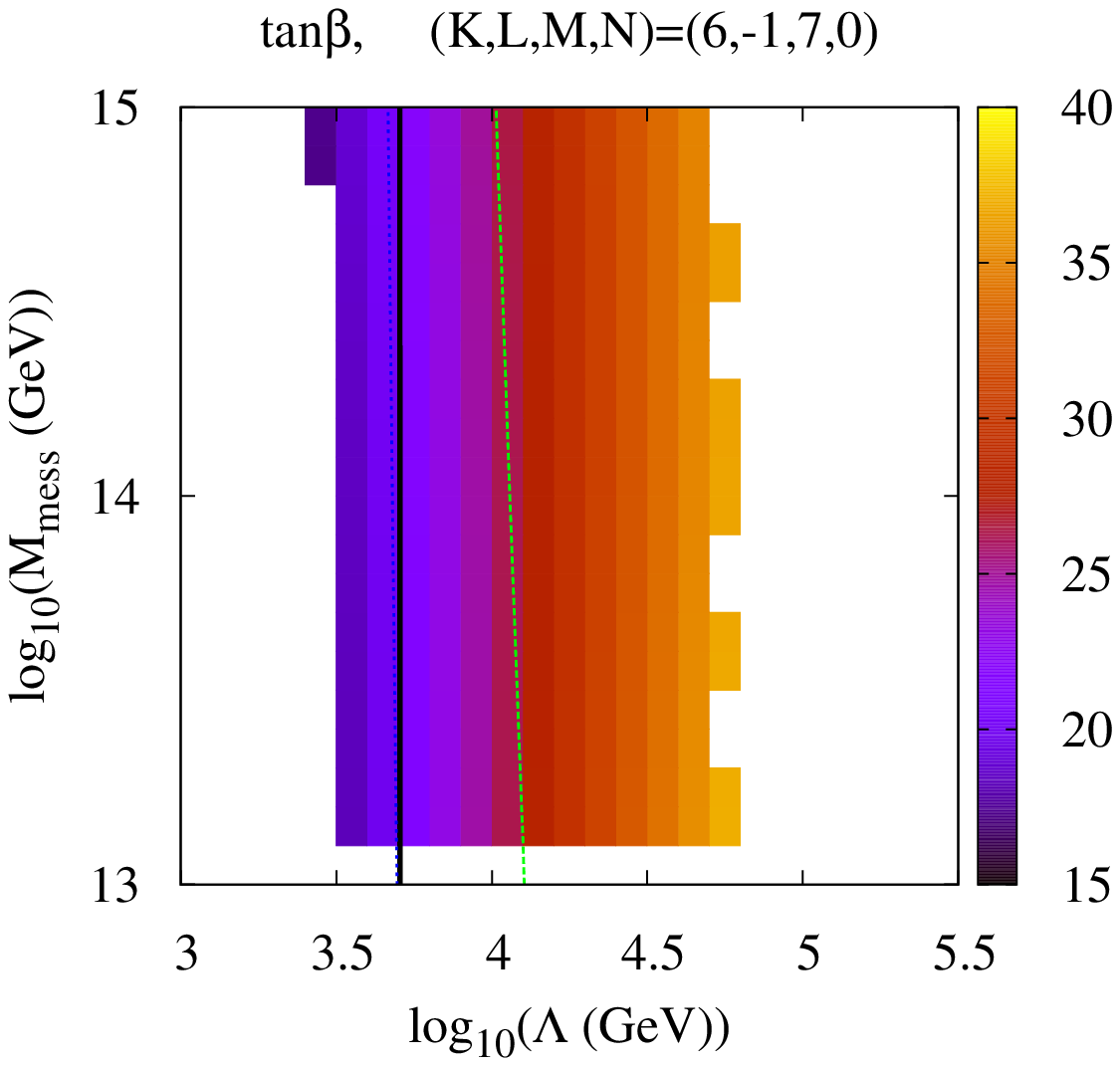}{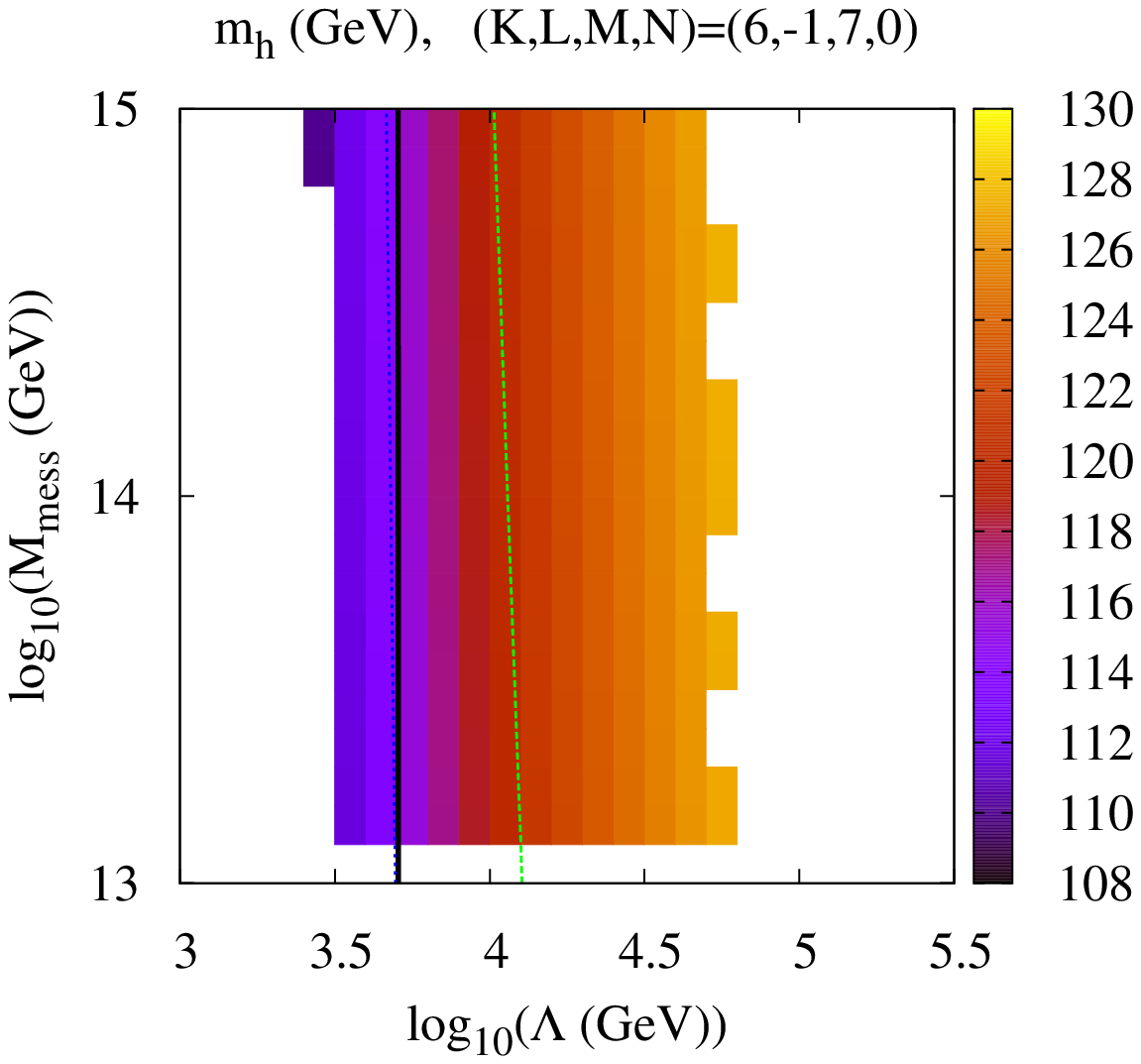}
\caption{The left-hand plot (a) shows the viable parameter range in the $\log_{10}(M_{\rm Mess})-\log_{10}(\Lambda)$ plane for the flux choice $(6,-1,7,0)$, i.e. $\hbox{NonUniv}_{L=-1}^{\rm{LargeFlux}}$. It is coloured according to the values of $\tan\beta$ obtained at each viable point. The right hand plot (b) shows the mass of the lightest Higgs boson $h^0$. In both plots the blue dashed contour line is the LEP bound on the Higgs mass, $m_h=114$~GeV, the black line represents $m_{\tilde{g}}=1$~TeV and the green dashed line is the CMS limit on stable charged particles, $m_{\tilde{\tau}}=293$~GeV.}

\label{fig:tanbplot2}
\end{center}
\end{figure}

\begin{figure}
\begin{center}
\epsfig{file=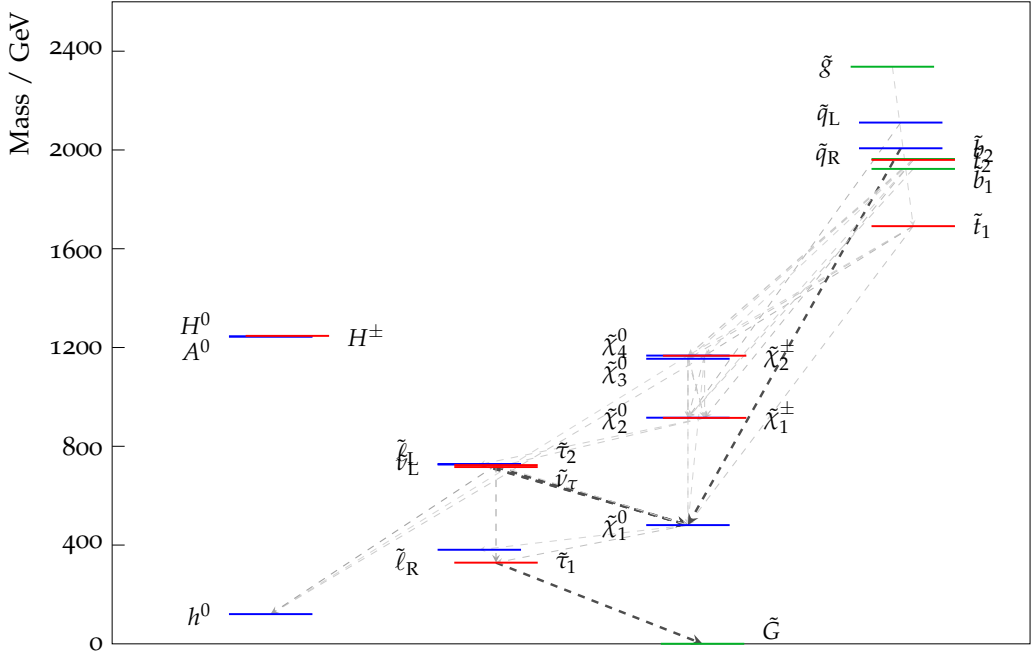,width=0.8\textwidth}
\caption{This figure shows the spectrum of a large flux benchmark point with $(K,L,M,N)= (6,-1,7,0)$, $M_{\rm Mess} = 10^{14}$~GeV and $\Lambda= 1.3\times 10^{4}$~GeV. The corresponding value of $\tan\beta$ is 28.}
\label{fig:LargeFluxSpec}
\end{center}
\end{figure}

Since the stau is stable due to the high messenger scale, it will appear in the detectors as a slow moving heavy muon-like particle, and the spectrum reconstruction prospects in this case are very good. For instance, the momentum and velocity of the stau can be measured in the inner tracker and muon calorimeter as it leaves the detector allowing sub-GeV accuracy in the reconstruction of the stau mass. Scenarios with long lived staus have recently been investigated in~\cite{Feng:2009bd,Ito:2009xy,Heckman:2010xz}. 
Using the methods of those papers it should be possible to reconstruct the masses of the right-handed sleptons to sub-GeV precision and  the two lightest neutralinos to GeV accuracy assuming $100fb^{-1}$ of data taken at $\sqrt{s}=14$~TeV. However, these papers assume that the mass of the lightest stau is around $\approx 150$~GeV. Since then the CMS collaboration have released the results of a search for such heavy stable charged particles~\cite{CMS-PAS-EXO-11-022}, which reports a new lower bound on the mass of the stable stau of 293~GeV corresponding to $1.09fb^{-1}$ of integrated luminosity. {While there is some unquantified model dependence associated with this precise value, it should still provide a good estimate of where the limit is likely to like in our model.}
The prospects for the re-construction of the {{spectrum must }}therefore be re-evaluated in detail. We note here that the new lower bound from CMS is approximately twice the assumed mass from the studies cited above. In a single scale model (such as mGMSB and our F-theory GUT), doubling the mass of the NLSP will nearly double the masses of all the sparticles in the spectrum. If doubling the masses decreases the cross-section by a factor of four, at least 400$fb^{-1}$ at 14~TeV will be needed to attain such precision.

Figure~\ref{fig:tanbplot2} shows the (a) values of $\tan\beta$ and (b) $m_h$ for this choice of fluxes. Also shown are the LEP lower bound on the Higgs mass and 1~TeV gluino mass contours as in the small flux case. The green dashed line is the CMS bound on the mass of the stau, 293~GeV. This is clearly the strongest constraint in the large flux regime, which leads to $\tan\beta > 30$ and $m_h > 120$~GeV.

Figure \ref{fig:LargeFluxSpec} shows a sample spectrum from this region of parameter space with $M_{\rm Mess}= 10^{14}$~GeV, $\Lambda = 1.3\times 10^4$~GeV and $\tan\beta = 28$. This point has been chosen so as not to be in conflict with the CMS heavy stable charged particle search discussed above, and the mass of the stau NLSP is 328~GeV. This is quite a strong constraint, which leads to the strongly interacting superpartners having masses around 2~TeV.
Due to both the large flux values (i.e. messenger index) and moderately high $\tan\beta$ the lightest neutralino with mass $481$~GeV is 152~GeV heavier than the stau NLSP. 
Measurements of the NLSP and $\tilde{\mu}_R$ and $\tilde{e}_R$ masses should provide a good way to constrain the value of $\tan\beta$ in a global parameter fit. While the mass splitting between the NLSP and the nearly-degenerate lightest smuon and selectron is $\sim 50$~GeV, large enough that the SM decay products from $\tilde{l}^{\pm} \to l^{\pm} \tau^{\mp} \tilde{\tau}^{\pm}$ should be hard enough to see, the $\tilde{\mu}_R-\tilde{e}_R$ splitting is only $\sim 150$~MeV. The left-handed slepton masses lie in between the lightest and next-to-lightest neutralino masses.

In gauge mediated theories the LSP is the gravitino, to which the NLSP will always eventually decay. The decay length of the NLSP is given by
\begin{equation}
\label{eq-withk}
L_{decay} = \frac{1}{\kappa} \left( \frac{100\mbox{GeV} }{m_{NLSP}} \right)^5
\left( \frac{\sqrt{\Lambda_G M_{\rm Mess}} }{100\mbox{TeV} } \right)^4 \times 10^{-4} {\rm m}\,,
\end{equation}
which can be found in~\cite{Giudice:1998bp}. The decay of the NLSP does not depend on the number of messengers present (or equivalently on the flux parameters in our model). Therefore the large flux regime $(K,L,M,N)=(6,-1,7,0)$, which requires smaller values of $\Lambda$ has a smaller NLSP decay length than the small-flux regime with $(K,M,L,N) = (2,-1,0,0)$. This distinction is somewhat pedantic however, as in neither case will the decay of the NLSP be observable in the detector. We show in Figure~\ref{fig:decaylength} the decay lengths of the NLSPs in these two scenarios, red for the small flux and blue for the large flux. In the small flux regime the decay length is always bigger than $10^{10}$ metres. There is a visible kink in the small flux points where the identity of the NLSP changes from neutralino to stau, which is associated with a change in the value of $\kappa$. In the large flux regime the minimum value of the decay length is just under $10^8$ metres.
\begin{figure}
\begin{center}
\epsfig{file=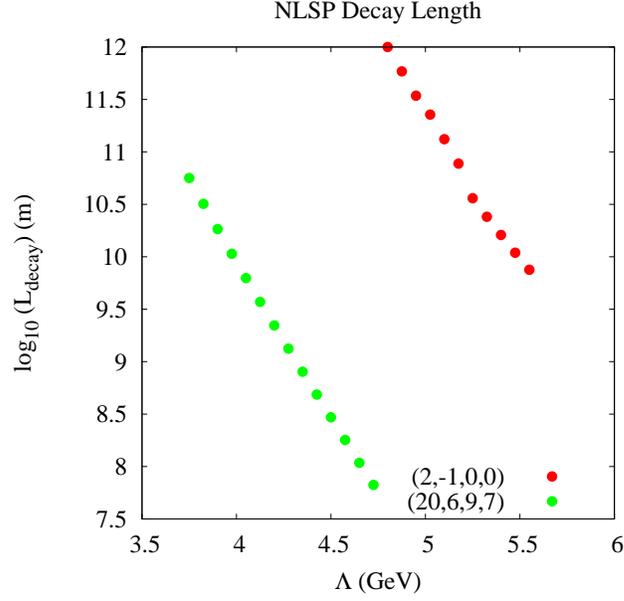,width=0.7\textwidth}
\caption{This plot shows the decay length in metres of the NLSP into the gravitino for the low flux (2,-1,0,0) (red circles) and large flux (6,-1,7,0)  (blue circles) scenarios, against the mass parameter $\Lambda$.}
\label{fig:decaylength}
\end{center}
\end{figure}

\subsubsection{$\hbox{NonUniv}_{N-L=2}$ and $\hbox{NonUniv}_{N-L=-2}$}

Figure~\ref{fig:0-210Spec} shows an example spectrum of a benchmark point with $(K,L,M,N)=(0,-2,1,0)$, $\Lambda= 9\times 10^4$~GeV and messenger scale $10^{14}$~GeV. The non-universalities in this model increase the bino-wino splitting and decrease the bino-gluino splitting. The electroweak symmetry breaking conditions lead to a large value of $\tan\beta$ (46) which in turn leads to stau NLSP due to large mixing proportional to $\tan\beta$. This also leads to significant splitting of the stops and sbottoms. We do not present parameter space scans for these last two scenarios since they are similar in content and form to those already presented for Models 1 and 2.
\begin{figure}
\begin{center}
\epsfig{file=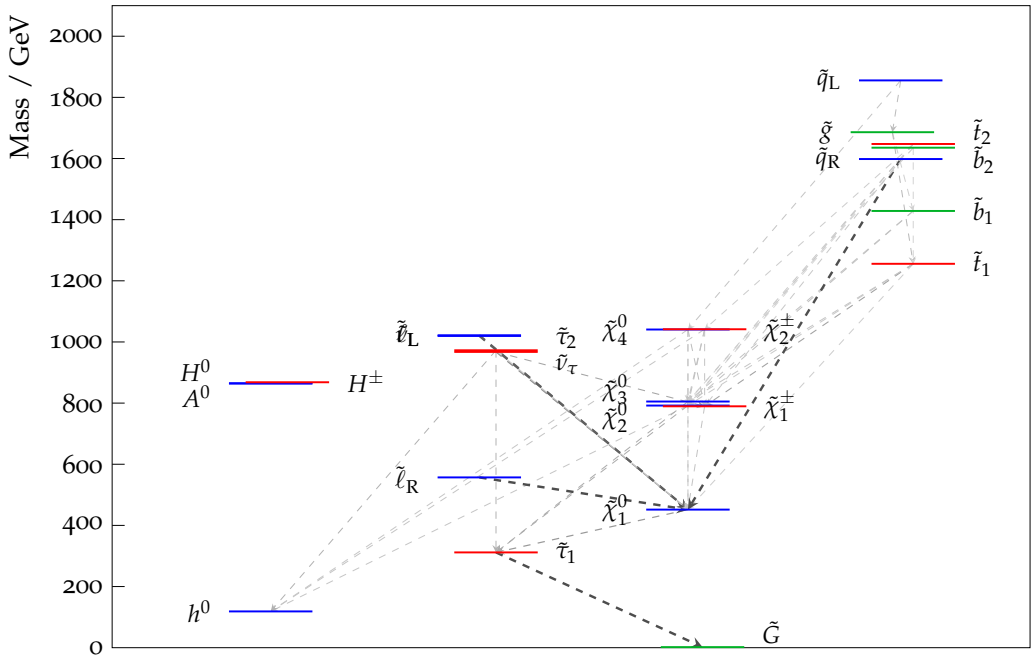,width=0.8\textwidth}
\caption{This figure shows the spectrum of $\hbox{NonUniv}_{N-L=2}$, $(K,L,M,N)=(0,-2,1,0)$ with $M_{\rm Mess} = 10^{14}$~GeV and $\Lambda= 8\times 10^{4}$~GeV.}
\label{fig:0-210Spec}
\end{center}
\end{figure}

We finally discuss the last benchmark point, which has $(K,L,M,N)=(3,2,0,0)$, $\Lambda=7\times 10^4$~GeV and messenger scale $M_{\rm Mess}=10^{14}$~GeV. 
This point is notable for its compressed slepton and light chargino spectrum. In models with gaugino universality the ratio of the bino to wino masses at the weak scale is approximately $1:2$. The flux induced non-universalities decrease this by a factor of 4/9, so that the weakly interacting part of the spectrum is extremely degenerate. There is only a 7~GeV splitting between the bino-like NLSP and the wino-like $\chi_2^0$ and $\chi_1^{\pm}$. The value of $\tan\beta$ at this point is 7, so that third-family mixing is also small, adding to the degeneracy.
On the other hand, the coloured part of the spectrum is over sixty percent heavier than in the universal case. This spectrum realises an extreme splitting between the coloured and uncoloured parts of the spectrum. However, this should not lead to any unusual signatures. Jets plus missing energy searches will still be sensitive probes of the existence of SUSY. However, it will be difficult to extract information about the leptonic part of the spectrum, due to the softness of any leptons emitted in long decay chains. 
\begin{figure}
\begin{center}
\epsfig{file=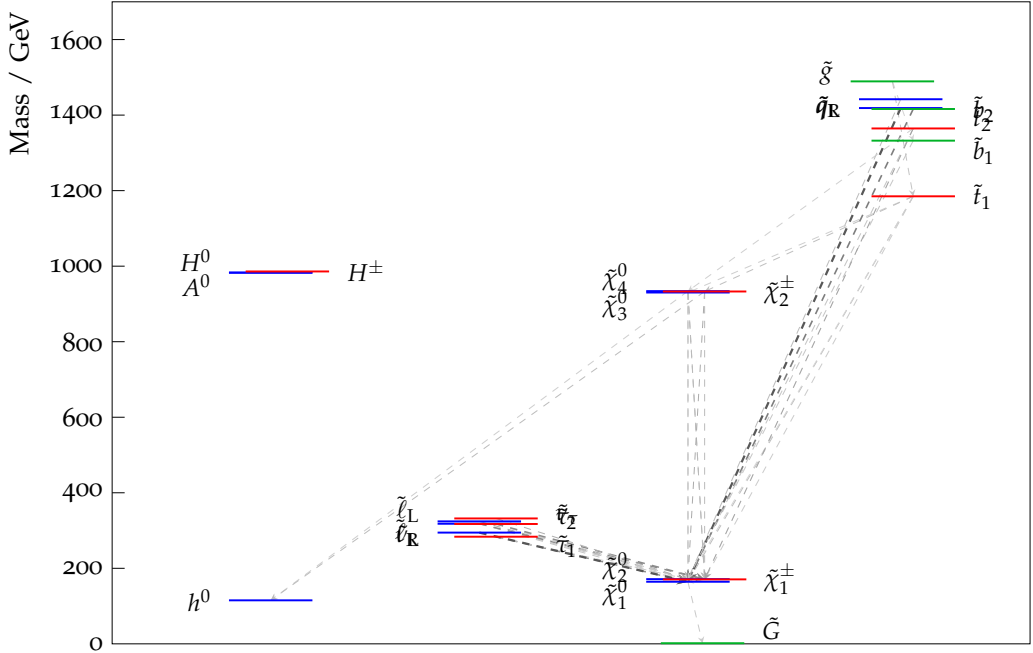,width=0.8\textwidth}
\caption{This figure shows the spectrum of $\hbox{NonUniv}_{N-L=-2}$ with $M_{\rm Mess} = 10^{14}$~GeV and $\Lambda= 7\times 10^{4}$~GeV.}
\label{fig:4200Spec}
\end{center}
\end{figure}

A summary of the identities of the NLSP (and other light sparticles) is shown in Figure~\ref{fig:BMNLSP}, showing four different patterns of light particles. 
{{In benchmarks 1 and 4 the stau is the NLSP, with the neutralino as the next to lightest particle. In model 2 the neutralino is the NLSP with the stau coming after that. In model 3 due to the compression of the spectrum the neutralino is NLSP and is nearly degenerate with the lightest chargino and wino.}}
\begin{figure}
\begin{center}
\epsfig{file=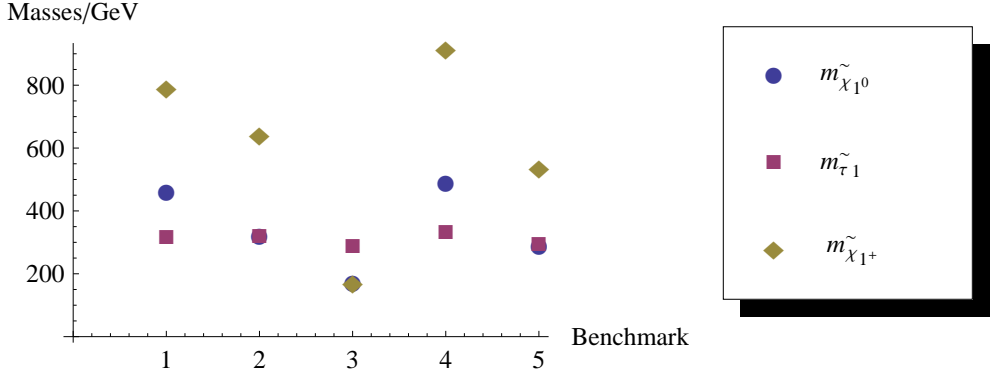,width=0.8\textwidth}
\caption{NLSPs for four Benchmark models, $\hbox{NonUniv}_{L=-2}$, $\hbox{NonUniv}_{L=-1}^{\rm{SmallFlux}}$,  $\hbox{NonUniv}_{L=2}$,  $\hbox{NonUniv}_{L=-1}^{\rm{LargeFlux}}$ and the small flux mGMSB lookalike. }
\label{fig:BMNLSP}
\end{center}
\end{figure}

\subsection{Decay Channels and Production Cross-Sections}

We now turn to the production and decay channels of the benchmark models. We consider first the production cross-sections for the coloured sparticles, which are expected to  dominate the total SUSY cross-section at the LHC. Figures~\ref{fig:BMLHC7} and~\ref{fig:BMLHC14} show the cross-sections for the main coloured production processes for the four benchmark points at the LHC with 7~TeV and 14~TeV respectively. 

At 7~TeV coloured production is dominated by $\tilde{q}\tilde{g}$ and $\tilde{q}\tilde{q}$. The cross-sections are quite small due to the masses of the squarks and gluino being between 1.5 and 2~TeV for the benchmark points. The largest total cross-section is in Benchmark 2, but even that is only 4.3fb. For Benchmark 4 the cross-section is miniscule, at 6.7 nanobarns. The two models with low cross-sections, $\hbox{NonUniv}_{L=-2}$ and $\hbox{NonUniv}^{\rm{LargeFlux}}_{L=-1}$ both have stau NLSP. In that case one sets limits on the stau mass under the assumption of only weak production processes. In general the prospects for the observation of the benchmark models at LHC7 are not very good. However, this is also partly a reflection of the the fact that they have been chosen to evade the constraints set in the limits set using $1fb^{-1}$ of data and reported at the summer 2011 conferences. 
\begin{figure}
\begin{center}
\epsfig{file=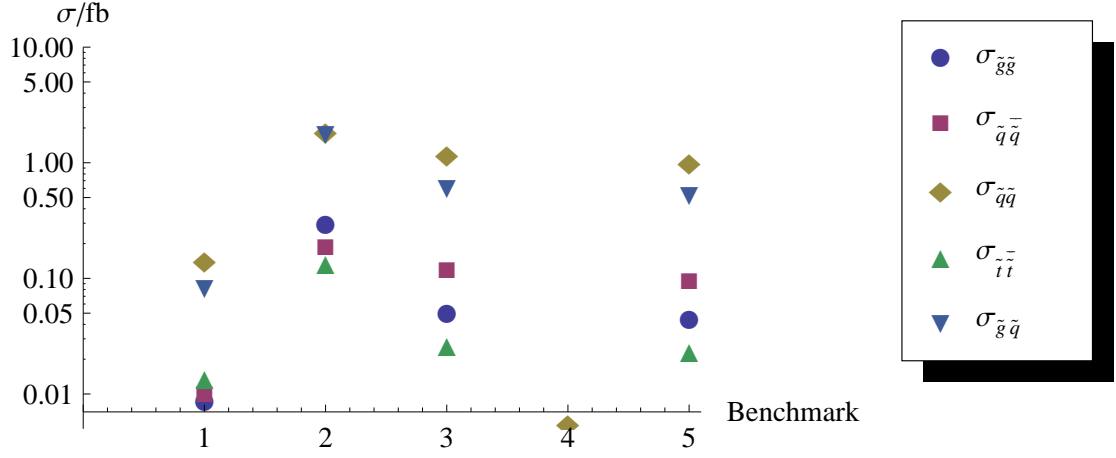,width=0.9\textwidth}
\caption{Production cross sections for the benchmarks $\hbox{NonUniv}_{L=-2}$, $\hbox{NonUniv}_{L=-1}^{\rm{SmallFlux}}$,  $\hbox{NonUniv}_{L=2}$,  $\hbox{NonUniv}_{L=-1}^{\rm{LargeFlux}}$ and small flux lookalike at 7~TeV. }
\label{fig:BMLHC7}
\end{center}
\end{figure}

At 14~TeV the situation is somewhat different, as the production is now dominated by $\tilde{g}\tilde{g}$ which was suppressed at 7~TeV, due to the large gluon component in the parton distribution functions at high energies. This difference in dominant production mechanisms leads to slightly different decay topologies predominating at each energy. At 7~TeV, the larger number of squarks should lead to more two and three jet events, plus missing energy (assuming that there is sufficient integrated luminosity to observe these events). The gluinos produced at 14~TeV will decay either through a three-body decay, or a two body decay via a squark. In either case, we expect 4 jets plus missing energy. Since this is due mainly to the PDFs, it is unlikely to serve as a robust signal of our particular model, and it is shared with the mGMSB lookalike point, Model 5.
\begin{figure}
\begin{center}
\epsfig{file=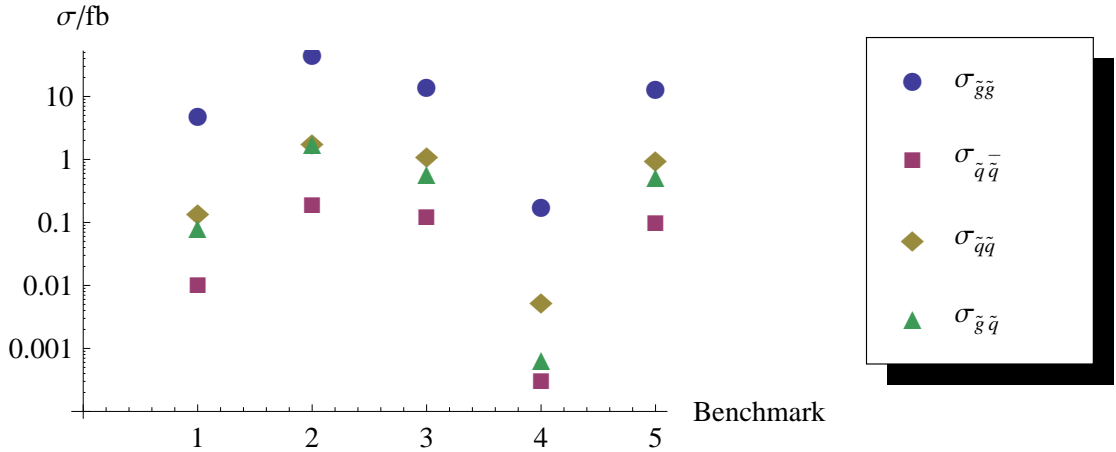,width=0.9\textwidth}
\caption{Production cross sections for the benchmarks $\hbox{NonUniv}_{L=-2}$, $\hbox{NonUniv}_{L=-1}^{\rm{SmallFlux}}$,  $\hbox{NonUniv}_{L=2}$,  $\hbox{NonUniv}_{L=-1}^{\rm{LargeFlux}}$ and small flux lookalike at 14~TeV. }
\label{fig:BMLHC14}
\end{center}
\end{figure}

We now turn to the decays of the gluino, since its production will dominate at LHC14 and the prospects at LHC7 are unimpressive. To calculate these branching ratios we have used SUSYHIT~\cite{Muhlleitner:2003vg}. The branching ratios and main decay channels of the gluino are shown for each of the benchmark points in Fig.\ref{fig:BMNLSP}. 
In models 1 and 2 the left-handed squarks are heavier than the gluino, while for models 3 and 4 the gluino is the heaviest sparticle. 
  In all four models the channel with the highest branching ratio is gluino decay to stop-top, with the total branching ratio to stop-top varying from 76\% in Model 1 to 22\% in Model 4. The stop then either decays to a top and a neutralino, or a bottom quark and chargino. This is a similar situation to the F-theory GUTs studied in~\cite{Heckman:2009bi} and as in that case quadruple top production should be an observable signal~\cite{Acharya:2009gb}.  After decay to stop-top the next most common decay mode is to sbottom-bottom, which varies from 12\% in Model 2 to 30\% in Model 3, leading to extensive heavy flavour production in all benchmark points.

\begin{figure}
\begin{center}
\epsfig{file=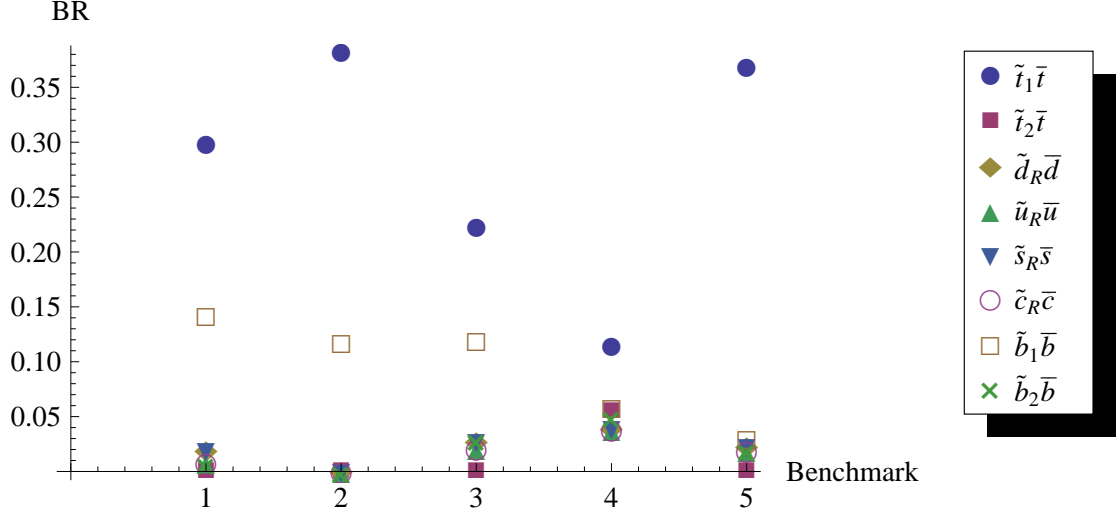,width=0.9\textwidth}
\caption{Gluino branching ratios for the benchmarks $\hbox{NonUniv}_{L=-2}$, $\hbox{NonUniv}_{L=-1}^{\rm{SmallFlux}}$,  $\hbox{NonUniv}_{L=2}$,  $\hbox{NonUniv}_{L=-1}^{\rm{LargeFlux}}$ and the mGMSB lookalike for  $\hbox{NonUniv}_{L=-1}^{\rm{SmallFlux}}$.}
\label{fig:BMgluinoBR}
\end{center}
\end{figure}

\subsection{Distinguishing from lookalike}

The flux-induced non-universalities lead to calculable deviations from the mass relations of minimal gauge mediation. In some limits, such as the large flux regime, the mass relations in F-theory models asymptotically approach the mGMSB relations. How measurable are the deviations in the small-flux regime? In principle this should be answered on a model-by-model basis for each choice of fluxes. Here we argue that for one of our benchmark points  the imprint of the non-universalities is such that precision kinematic edge measurements~\cite{Barr:2010zj} will allow a discrimination between the two models.

We will compare the $\hbox{NonUniv}_{L=-1}^{\rm{SmallFlux}}$ benchmark with the mGMSB lookalike whose spectrum is shown in Figure~\ref{fig:mGMSBspec1}.  While there are differences in the ratios between the gaugino and sfermion masses for these points, to leading order we would expect the phenomenology of both points to be broadly similar, due to the similarity of the strongly interacting spectrum in both cases. Further, we have chosen the value of the messenger index so that the decay channels (in particular the fact that the gluino decays to squarks, rather than gauginos) and branching ratios are similar between our model point and the mGMSB one.

We will focus on the `golden channel'  $\tilde{q}_L \to \chi_2^0 q \to \tilde{l}_R^{\pm} l^{\mp} q \to \chi^0_1 l^{\pm} l ^{\mp} q$. 
Of particular relevance is the study~\cite{Allanach:2000kt}\footnote{More recent work includes~\cite{Allanach:2011ya} which demonstrates that mGMSB cannot always be discriminated from the CMSSM.}, which focusses on model discrimination between mSUGRA and a non-universal string-derived model. Since our F-theory GUT predicts a high messenger scale and hence stability of the NLSP on collider timescales, this analysis should be applicable in our scenario as well. 
The following edges and thresholds are present in both the mGMSB and F-theory points. First, the $l^+ l^-$ edge from the dilepton invariant mass spectrum caused by $\chi^0_2 \to \tilde{l}^{\pm}l^{\mp} \to \chi^0_1 l^{\pm} l^{\mp}$. We also consider the $l^+ l^- q$ edge and threshold, the $l^{\pm}q$ high- and low- edges, and finally the $M_{T2}$ edge, which can be used to constrain the lightest slepton-neutralino mass difference. Analytic expressions for all these variables can be found in~\cite{Allanach:2000kt}.
 With $100fb^{-1}$ at $\sqrt{s}=14$~TeV, the authors of~\cite{Allanach:2000kt} achieve an accuracy in reconstruction of $1-2\%$ in $m_{ll}$, and $1-4\%$ in the variables involving the quarks. In our case, the masses of the benchmark points are somewhat higher, and hence a larger amount of integrated luminosity is likely to be required for a similar precision. However, $200fb^{-1}$ still only represents two years running at high luminosity, although it will be some time before such levels are achieved. With this proviso, we show in Table~\ref{tab:kinematics} the values of the kinematic edges for the F-theory and mGMSB model points.

We see that the lookalike point is easily distinguishable from the the $\mbox{NonUniv}^{\rm{SmallFlux}}_{L=-1}$ point. We have extensively searched in the mGMSB parameter space with $N_5 =2$ and $3$ for points that are significantly better than this, but were unable to find any. For instance, if one adjusts the mGMSB parameters to achieve $m_{ll}^{GMSB} \approx 330$~GeV, then the coloured particle masses are above 2~TeV, so that the total SUSY cross-section becomes very different. It is this tension which prevents the F-GUT being indistinguishable from mGMSB: if a point is chosen to match the coloured cross-section then the spectrum compression in the weak sector leads to bad agreement between the kinematic edges. On the other hand, choosing an mGMSB point with similar kinematic edge values means that the total cross-section will be much lower than the F-theory benchmark point. 
\begin{table}
\begin{center}
\begin{tabular}{|c|c|c|c|c|c|} \hline
Model & $m_{ll}$ & $m_{llq}^{\mbox{edge}}$ & $m_{llq}^{\mbox{thr}}$ & $m_{lq}^{\mbox{high}}$ & $m_{lq}^{\mbox{low}}$ \\ \hline
mGMSB & 138.7 & 1126 & 306 & 1102 & 396 \\ \hline
F-GUT & 330.2 & 1011 & 550 & 856 & 688  \\ \hline
\end{tabular}
\caption{ This table shows the values of the kinematic invariants discussed in the text for the F-theory low-flux benchmark point $\mbox{NonUniv}^{\rm{SmallFlux}}_{L=-1}$ with $(K,L,M) = (2,-1,0)$ and the mGMSB lookalike point, whose spectra are shown in Figures~\ref{fig:M1N1Spec} and~\ref{fig:mGMSBspec1} respectively.}
\label{tab:kinematics}
\end{center}
\end{table}

Based on this discussion and comparison with the results of~\cite{Allanach:2000kt}, we believe it should be possible to discriminate between our model and mGMSB when the non-universalities are at their largest.
We are aware that we have not shown systematically that distinguishing between the two models is always possible. That would involve detailed simulations, event generation and reconstruction that is beyond the scope of this article. However, we hope to return to this issue in the future.

\section*{Acknowledgements}

{We thank K.~Bobkov, P.~Kumar, N.~Saulina, and M.~Wijnholt for discussions.}
We would like to thank
the organizers of the MPI Munich workshop "GUTs and Strings" for providing a stimulating
research environment where this collaboration began.
MJD thanks CERN for hospitality during the completion of this work. SS-N thanks the Caltech theory group and the KITP for hospitality during the course of this work.


\appendix

\section{Simple Corrections to RG}
\label{app:RG}
In this appendix we determine the corrections to the running of the gauge couplings that are due to NLO effects and low-scale thresholds. All these effects are standard MSSM corrections. 


\subsection{2-loop Running}

The $\beta$ functions for gauge couplings are expressed to 2-loop order as
\begin{equation}\beta_i(g) =b_i\left(\frac{g_i^3}{16\pi^2}\right) + \sum_{j=1}^3 b_{ij}\left(\frac{g_i^3 g_j^2}{(16\pi^2)^2}\right) \,.\end{equation}
In order to be completely clear about normalizations, the explicit 2-loop RGE that we obtain is
\begin{equation}\mu\frac{\partial}{\partial\mu}\alpha_i^{-1} = -\frac{b_i}{2\pi}-\sum_j \frac{b_{ij}\alpha_j}{8\pi^2} \,.\label{RGE}\end{equation}
To include next-to-leading order corrections from the 2-loop $\beta$ function, we replace the $\alpha_j$ on the right hand side of \eqref{RGE} by the 1-loop expression
\begin{equation}\alpha_j^{(1-loop)}(\mu) = \frac{\alpha_{U}}{1 + \frac{b_j\alpha_{U}}{2\pi}\ln\left(\frac{M_U}{\mu}\right)}\,.\end{equation}
Plugging this into \eqref{RGE} and integrating yields
\begin{equation}\alpha_i^{-1}(m) = \alpha_i^{-1}(m)^{(1-loop)} + \sum_{j=1}^3 \frac{b_{ij}}{4\pi b_j}\ln\left[1 + b_j\left(\frac{\alpha_{U}}{2\pi}\right) \ln(M_{U}/m)\right] \,.
\end{equation}
The resulting 2-loop correction parameters $\delta_i^{(2-loop)}$ are read off as
\begin{equation}\delta_i^{(2-loop)} = \sum_{j=1}^3 \frac{b_{ij}}{4\pi b_j}\ln\left[1+ b_j\left(\frac{\alpha_{U}^{(0)}}{2\pi}\right)\ln\left(\frac{M_{U}^{(0)}}{m_Z}\right)\right] \,.
\end{equation}
To get a leading order correction, we insert the 1-loop values of $M_U$ and $\alpha_U$, namely the expressions \eqref{alphaMU1loop}.  This leads to{\footnote{This agrees with the expression in equation (42) of \cite{Alciati:2005ur} once we set $b_1=33/5$ and $b_2=1$ up to a factor of 4 that arises from a difference in our convention for the $b_{ij}$.}}
\begin{equation}\delta_i^{(2-loop)} = \sum_{j=1}^3 \frac{b_{ij}}{4\pi b_j}\ln\left[1 + b_j\left(\frac{3-8\sin^2\theta_W(m_Z)}{5b_1\sin^2\theta_W(m_Z) - 3b_2\cos^2\theta_W(m_Z)}\right)\right] \,.\end{equation}
To actually evaluate this correction, we need the 2-loop coefficients.  These are well-known and take the form\cite{Einhorn:1981sx,Jones:1981we}
\begin{equation}\begin{split}b_{ij} &= \begin{pmatrix}0 & 0 & 0 \\ 0 & -24 & 0 \\ 0 & 0 & -54\end{pmatrix} + N_{gen}\begin{pmatrix}38/15 & 6/5 & 88/15 \\ 2/5 & 14 & 8 \\ 11/15 & 3 & 68/3\end{pmatrix} + N_{Higgs}\begin{pmatrix}9/50 & 9/10 & 0 \\ 3/10 & 7/2 & 0 \\ 0 & 0 & 0\end{pmatrix} \\
&= \begin{pmatrix}199/25 & 27/5 & 88/5 \\ 9/5 & 25 & 24 \\ 11/5 & 9 & 14\end{pmatrix}
\end{split}\end{equation}
where we have set the number of generations to $N_{gen}=3$ and the number of Higgs doublets $N_{Higgs}=2$ in the last line.

Using these, the $\delta_i^{(2-loop)}$ corrections are given by
\begin{equation}\delta_i^{(2-loop)} = \begin{pmatrix} 0.6570 \\ 1.074 \\ 0.5517 \end{pmatrix}\,,\end{equation}
whose contribution to $\Delta$ is
\begin{equation}\Delta^{(2-loop)} = -0.8197 \,.\end{equation}
Note that this is rather large in comparison to the deviation of $\Delta^{(1-loop)}$ from $\Delta^{(\text{ideal})}$ \eqref{Deltaideal}.  Let us also write out the 2-loop shifts in $M_U$ and $\alpha_U^{-1}$
\begin{equation}\begin{split}\delta\left(\frac{1}{2\pi}\ln\frac{M_U}{m_Z}\right)^{(2-loop)} & = \frac{\delta_2-\delta_1}{b_1-b_2} \\
& = 0.07446 \\
\delta\left(\alpha_U^{-1}\right)^{(2-loop)} &= \frac{b_2\delta_1 - b_1\delta_2}{b_1-b_2} \\
&= -1.148 \,.
\end{split}\end{equation}


\subsection{Low Scale Thresholds}

Next, we look to threshold corrections from the low end of the RG flow.  Ordinary MSSM running assumes that all components of all supermultiplets run from $m_Z$ up to $M_{\rm GUT}$.  This is not the case, however, because superpartners do not kick in until their masses, which are all larger than $m_Z$.  This means that we have to subtract off the running of superpartners from their masses, $m_j$ down to $m_Z$, leading to the correction
\begin{equation}\delta_i^{(\text{low thresh})} = -\frac{1}{2\pi}\sum_j b_i(j)\ln\left(\frac{m_j}{m_Z}\right) \label{eq:lowscaledelta} \,,\end{equation}
where $j$ runs over the SUSY particle spectrum and any Higgs fields floating around and $b_i(j)$ denotes the corresponding contributions to the $\beta$ functions. 
It is possible to define an effective threshold scale $m_{SUSY}$ which would produce the same threshold corrections as (\ref{eq:lowscaledelta}) in the case where each type of supersymmetric particle has the same mass.  It can be shown that this scale is~\cite{Carena:1993ag,Yamada:1992kv}
\begin{equation}
m_{SUSY} = m_{\tilde{H}} \left( \frac{m_{\tilde{W}}}{m_{\tilde{g}}} \right)^{28/19}
\left[  \left( \frac{m_{\tilde{l}}}{m_{\tilde{q}}} \right)^ {3/19}  \left( \frac{m_{H}}{m_{\tilde{H}}} \right)^{3/19} 
\left( \frac{m_{\tilde{W}}}{m_{\tilde{H}}} \right)^{4/19}     \right] 
\,, \label{msusydef}\end{equation}
which holds for sparticle masses larger than $m_Z$. In the case of independent universal scalar and gaugino masses at the GUT scale this approximately reduces to $m_{SUSY} \sim |\mu|/7$.

Using $m_{SUSY}$ to include low scale thresholds, the proper 1-loop running is expressed as 
\begin{equation}\frac{1}{\alpha_i}(m_Z) = \frac{1}{\alpha_i(M_{\rm GUT})} + \frac{b_i^{(SM)}}{2\pi}\ln\left(\frac{m_{SUSY}}{m_Z}\right) + \frac{b_i^{(MSSM)}}{2\pi}\ln\left(\frac{M_{\rm GUT}}{m_{SUSY}}\right) \,.\end{equation}
This means that we can encapsulate this effect by the correction
\begin{equation}\delta_i^{(\text{low thresh})} = -\frac{b_i^{(MSSM)}-b_i^{(SM)}}{2\pi}\ln\left(\frac{m_{SUSY}}{m_Z}\right) \,,\end{equation}
where $b_i^{(MSSM)}$ are the MSSM $\beta$ function coefficients \eqref{MSSMbeta} and $b_i^{(SM)}$ are the Standard Model $\beta$ function coefficients.  The latter are given by
\begin{equation}b_j^{(SM)} = \begin{pmatrix}41/10 \\ -19/6 \\ -7\end{pmatrix} \,.\end{equation}
With this approximation, we get
\begin{equation}\delta^{(\text{Light Thresh})} = \frac{1}{2\pi}\ln\left(\frac{m_{SUSY}}{m_Z}\right)\times \begin{pmatrix}-5/2 \\ -25/6 \\ -4\end{pmatrix} \,,\end{equation}
whose contribution to $\Delta$ is
\begin{equation}\Delta^{(\text{Light Thresh})} = \frac{19}{28\pi}\ln\left(\frac{m_{{SUSY}}}{m_Z}\right) \,.\end{equation}
In addition to this, $M_U$ and $\alpha_U$ are shifted according to
\begin{equation}\begin{split}\delta\left(\frac{1}{2\pi}\ln\frac{M_U}{m_Z}\right)^{(\text{Light Thresh})} &= \frac{\delta_2-\delta_1}{b_1-b_2} \\
&= -\frac{25}{168\pi}\ln\frac{m_{{SUSY}}}{m_Z} \\
\delta\left(\alpha_U^{-1}\right)^{(\text{Light Thresh})} &= \frac{b_2\delta_1 - b_1\delta_2}{b_1-b_2} \\
&= \frac{125}{56\pi}\ln\frac{m_{{SUSY}}}{m_Z} \,.
\end{split}\end{equation}

\newpage


\bibliographystyle{JHEP}
\bibliography{FGUTbib}

\end{document}